\documentclass[10pt,preprint2]{aastex}

\newcommand{\be}{\begin{equation}}
\newcommand{\ee}{\end{equation}}

\newcommand{\dd}{{\rm d}}
\def\simlt{\lower.5ex\hbox{\ltsima}}
\def\gtsima{$\; \buildrel > \over \sim \;$}
\def\simgt{\lower.5ex\hbox{\gtsima}}

\def\simlt{\lower.5ex\hbox{\ltsima}}
\def\gtsima{$\; \buildrel > \over \sim \;$}
\def\simgt{\lower.5ex\hbox{\gtsima}}

\def\cm{{\rm\,cm}}

\def\ergcm2{\ {\rm erg~cm^{-2} }}
\def\ergscm2{\ {\rm erg~s^{-1}~cm^{-2} }}

\def\cm2s{\ cm^2 ~s^{-1} }

\def\s{\ifmmode \widetilde \else \~\fi}
\def\={\overline}

\def\spose#1{\hbox to 0pt{#1\hss}}

\def\lta{\mathrel{\spose{\lower 3pt\hbox{$\mathchar"218$}}
     \raise 2.0pt\hbox{$\mathchar"13C$}}}
\def\gta{\mathrel{\spose{\lower 3pt\hbox{$\mathchar"218$}}
     \raise 2.0pt\hbox{$\mathchar"13E$}}}
\def\mincir{\ \raise -2.truept\hbox{\rlap{\hbox{$\sim$}}\raise5.truept  
\hbox{$<$}\ }}                                                          %
\def\magcir{\ \raise -2.truept\hbox{\rlap{\hbox{$\sim$}}\raise5.truept  %
\hbox{$>$}\ }}                                                          %
\def\simlt{\ \raise -2.truept\hbox{\rlap{\hbox{$\sim$}}\raise5.truept   
\hbox{$<$}\ }}                                                          %
\def\simgt{\ \raise -2.truept\hbox{\rlap{\hbox{$\sim$}}\raise5.truept   %
\hbox{$>$}\ }}                                                          %
\def\newline{\par\noindent}

\def\simleq{\; \raise0.3ex\hbox{$<$\kern-0.75em \raise-1.1ex\hbox{$\sim$}}\; }
\def\simgeq{\; \raise0.3ex\hbox{$>$\kern-0.75em \raise-1.1ex\hbox{$\sim$}}\; }
\newcommand{\eV}{{\rm eV}}
\newcommand{\keV}{{\rm keV}}

\newcommand{\Mpc}{{\rm Mpc}}
\newcommand{\kpc}{{\rm kpc}}
\newcommand{\muG}{\mu{\rm G}}

\begin{document}

\title{Constrained Simulations of
the Magnetic Field in the Local Universe and the Propagation of
UHECRs}

\author{Klaus Dolag$^1$, Dario~Grasso$^2$, Volker~Springel$^3$ and
Igor Tkachev$^4$}

\altaffiltext{1}{Dipartimento di Astronomia, Universit\`a di
Padova, Padua, Italy}

\altaffiltext{2}{Scuola Normale Superiore and I.N.F.N., Pisa -
Italy}

\altaffiltext{3}{Max-Planck-Institut f\"ur Astrophysik, Garching,
Germany}

\altaffiltext{4}{CERN - Theory Division, Geneve, Switzerland}

\begin{abstract}
We use simulations of large-scale structure formation to study the build-up of
magnetic fields (MFs) in the intergalactic medium.  Our basic assumption is
that cosmological MFs grow in a magnetohydrodynamical (MHD) amplification
process driven by structure formation out of a magnetic seed field present at
high redshift. This approach is motivated by previous simulations of the MFs
in galaxy clusters which, under the same hypothesis that we adopt here,
succeeded in reproducing Faraday rotation measurements (RMs) in clusters of
galaxies.  Our $\Lambda$CDM initial conditions for the dark matter density
fluctuations have been statistically constrained by the observed large-scale
density field within a sphere of 110 Mpc around the Milky Way, based on the
IRAS 1.2-Jy all-sky redshift survey.  As a result, the positions and masses of
prominent galaxy clusters in our simulation coincide closely with their real
counterparts in the Local Universe.  We find excellent agreement between RMs
of our simulated galaxy clusters and observational data.  The improved
numerical resolution of our simulations compared to previous work also allows
us to study the MF in large-scale filaments, sheets and voids. By tracing the
propagation of ultra high energy (UHE) protons in the simulated MF we
construct full-sky maps of expected deflection angles of protons with arrival
energies $E=10^{20}~\eV$ and $ 4 \times 10^{19}~\eV$, respectively.
Accounting only for the structures within 110 Mpc, we find that strong
deflections are only produced if UHE protons cross galaxy clusters. The total
area on the sky covered by these structures is however very small. Over still
larger distances, multiple crossings of sheets and filaments may give rise to
noticeable deflections over a significant fraction of the sky; the exact amount
and angular distribution depends on the model adopted for the magnetic seed
field. Based on our results we argue that over a large fraction of the sky the
deflections are likely to remain smaller than the present experimental angular
sensitivity.  Therefore, we conclude that forthcoming air shower experiments
should be able to locate sources of UHE protons and shed more light on the
nature of cosmological MFs.
\end{abstract}


\section{Introduction}

In spite of decades of experimental and theoretical research on the
subject, the nature and origin of Ultra High Energy Cosmic Rays
(UHECRs) are still unknown. The experimental study of Extended Air
Showers (EAS) produced by the interaction of UHECRs with the
atmosphere provided, so far, only weak constraints on the composition
of the primary particles. One of the few established facts is that the
primaries of EAS are not standard-model neutrinos, since in this case
showers would develop too deep in the atmosphere.  Somewhat less
certain are the constraints on the allowed fractions of heavy nuclei
and photons.  The search of a possible asymmetry between vertical and
inclined showers made by the Haverah Park collaboration
\citep{HavPar..photons} allowed the conclusion that at energies above
$4 \times 10^{19}~\eV$ less than $50\%$ of EAS primaries are
photons. Concerning the fraction of nuclei, the analysis of the
inclined events recorded by the Fly-Eye shower detector
\citep{FlyEye..comp} seems to favor a light composition above
$10^{19}~\eV$. 
Similar results concerning photonic and heavy nuclei fractions were
obtained by AGASA \citep{Shinozaki:2003vv} using muon density in air
showers as a primary mass estimator. It is therefore plausible that a
significant component of UHECRs is composed of protons or light nuclei.

Besides determining the composition and energy spectrum of UHECR
primaries, one of the main goals of EAS experiments is to map the
angular distribution of cosmic rays. The angular resolution of these
experiments lies generally in the range $0.5-2$ degrees, and may thus in
principle allow the identification of UHECR sources with astrophysical
counterparts observed in some other channel (optical, radio, X-rays,
gamma-rays).  While the arrival directions of neutral particles would
directly point to the sources (gravitational lensing is negligible), 
protons and composite nuclei may be deflected by galactic and
Extragalactic Magnetic Fields (EGMFs). Note that at extremely high
energies, $E \simgeq 10^{20}~\eV$, UHE protons are not expected to be
significantly deflected by the {\em galactic} magnetic fields.  Even
at lower energies, $E \simgeq 4\times 10^{19}~\eV$, the turbulent
component of the MF in the Milky Way does not produce sizable
deflections, while the large-scale component gives rise to ordered
deflections, such that a reconstruction of the original arrival
directions may still be possible~\citep{Tinyakov:2001ir, Tinyakov:2003nu}. 
However, the very attractive perspective to do charged-particle astronomy 
with UHECRs might be spoiled by the presence of strong EGMFs.

So far, direct evidence for the presence of EGMFs has been found only
in galaxy clusters \citep[see e.g.][a short review of EGMF
observations will be also given in section
\ref{obs}]{Carilli.Taylor..2002}. The origin of these fields is still
unknown. One possibility is that the Intra Cluster Medium (ICM) has
been magnetized at low redshift by the ejecta of local sources. If
this is the case, EGMFs should be largely confined within galaxy
clusters and groups, so that significant deflections of charged UHECRs
would have to be expected only if they cross these structures.

Another possibility is that the seeds of the MFs observed in galaxy
clusters have originated at high redshift, before the clusters
collapsed and formed gravitationally bound systems. We will discuss
possible origins of the intra-cluster magnetic fields in more detail
in Section \ref{models}, and demonstrate in Section \ref{results1}
that this scenario -- which is the model studied in our work -- leads
to cluster MFs which reproduce many observational facts. Note that in
this case EGMFs should fill a large fraction of the volume of the
universe, so that they might induce considerable deflections of UHECRs
traveling over cosmological distances, provided the fields are strong
enough. Clearly this is an issue which has to be investigated
carefully.

Until recently, the possible effects of EGMFs on the propagation of UHECRs
have been investigated by constructing toy models for the structure of the
field. 
Typically, EGMFs were assumed to have a cellular structure, with a
Kolmogorov power spectrum at scales smaller than the cell size, and a uniform
correlation length and rms field strength. None of these assumptions, however,
is well supported by observational or theoretical arguments.  Indeed, due to
the large conductivity of the IGM, magnetic fields should follow the gas flow
and become concentrated and amplified in the high density regions.  The
assumption of a statistically homogeneous MF is hence not valid in a
structured Universe on scales of interest for the problem of UHECR
propagation.

In the last few years, several groups started to develop physically more
realistic models based on numerical simulations \citep{Ryu..1998, Sigl:2003ay,
Dolag:2003ft, Dolag:2003ra}, combining the magneto-hydrodynamics (MHD) of the
magnetized IGM with N-body simulations of the driving gravitational dynamics
of the dark matter. Barring this common general approach, the few simulations
performed so far differ however not only in their numerical methodology but
also in several important aspects of the model assumptions employed, as well
as in their final results.  The common problem faced by all of these attempts
is that one wants to predict the observationally unknown strength and geometry
of the magnetic field in low density environments (e.g. filaments and voids)
based on a calibration of the underlying model against observations of
magnetic fields in galaxy clusters.  This involves an extrapolation over
several orders of magnitude in density into a regime where there is no data.
Even having such uncertainties, it is nevertheless crucial to demonstrate that
the model, in the first place, correctly reproduces available observational
data on the cluster scale.

In a recent letter \citep{Dolag:2003ra}, we presented first
results of a novel simulation method for the magnetic structure of the Local
Universe, where we achieved a number of improvements compared with previous
works.  First, the initial conditions for the dark matter density fluctuations
were constrained to reproduce, to good approximation, the positions and masses
of the largest structures (galaxy clusters) observed around the Milky Way.
This allows us to remove the ambiguity in the choice of observer position that
exists in ordinary simulations, and to obtain the first simulated all-sky maps
of expected UHE proton deflections in the magnetic large-scale structure
around the Local Group. Since the simulation volume (a sphere with radius
$\sim 110\, \Mpc$ embedded in a box of $\sim 340\ \Mpc$ on a side) is much
larger than considered in previous work, even massive clusters are present in
our simulation, allowing for a more accurate comparison with
observations. Second, we used a Lagrangian simulation method with an adaptive
resolution scheme, which provides for better spatial resolution where it is
most needed, namely in the high density regions where the field becomes more
tangled. This helps in simulating RMs in galaxy clusters accurately, and hence
to establish a reliable baseline for comparing the MHD simulation results with
observations. By tracing UHE proton propagation in the simulated magnetic
structure obtained in this novel approach, we were then able to show that
significant deflections are only expected when galaxy clusters are
crossed. Since clusters cover only a very small fraction of the sky, UHECR
astronomy should therefore be, in principle, possible.

In this paper, we present a more detailed analysis of the predicted structure
of the magnetic field in galaxy clusters, filaments and voids, as well as UHE
proton propagation in the simulated magnetic web.  Our simulations are very
CPU demanding and so far we have made two runs with different initial magnetic
seed field. The results of the first run were reported in
\citet{Dolag:2003ra}.  The analysis of the present paper is mainly based on
results obtained in a second simulation, which has been improved in several
respects. In this new run, the comoving intensity of the seed field
\footnote{We define it as a value which the MF seed would have at present if
  only the cosmological redshift is taken into account. With this definition 
$B_0$ does not depend upon initial time.}
was chosen
to be $B_0 = 2 \times 10^{-12}$ G, a factor 5 smaller than in the previous
simulation.  This seed field strength leads to a more reasonable match with the
magnetic field strengths observed in galaxy clusters, whereas the first
simulation was pushing toward the upper limit of field strengths allowed by
the data. We changed the orientation of the uniform seed field in
order to test a possible dependence of our results on this aspect of the
initial conditions. We have also improved the MHD component of our code to
account for the back-reaction of the magnetic field on the ionized gas, which
in turn may affect the MF in the most dense regions. Our analysis takes into
account energy losses of UHE protons on cosmic microwave background (CMB)
photons, and we construct deflection maps of protons with an arrival energy of
$4\times 10^{19}~\eV$ and $10^{20}~\eV$.

This paper is structured as follows. In Section~2, we give an overview
of observations of magnetic fields and the possible origin of magnetic
fields in galaxy clusters, which provides the motivation for the model
we explore.  Section~3 describes the performed simulations, the
initial conditions used, and gives a brief description of the
simulation code used. Section~4 discusses the predictions of our
simulations for the strength and structure of magnetic fields in
galaxy clusters and relates these results to observations of Faraday
rotation and radio halos of clusters. Section~5 analyses the
simulation results for the magnetic field in low density environments
like filaments, sheets and voids. In Section~6, we discuss the
implications of the simulated magnetic field for the propagation of
UHE protons. Finally, we summarize our findings in Section~7.
In that section we also compare our approach with that
of Sigl et al. (2003,2004a)
and discuss the possible reasons of our discrepant results. 
Throughout this work we assume a background cosmology with
$H_0=0.7$, $\Omega_m=0.3$, $\Omega_\Lambda=0.7$, and a baryonic
fraction $\Omega_b/\Omega_m$ of 14\%.

\section{Observations of ICMFs and their possible origin}

\subsection{Observations}\label{obs}

Direct evidence of the presence of intra-cluster magnetic fields (ICMFs) is
provided by observations of extended radio halos in galaxy clusters. Their
radiation can only be due to synchrotron emission of relativistic electrons in
the ICM. The strength of the ICMFs can be estimated from the intensity of the
observed radio emission, either assuming the minimum energy condition, giving
$\left<B\right> \sim 0.1- 1~\muG$ \citep{1999dtrp.conf....3F,
1993ApJ...406..399G}, or by an independent determination of the density of
relativistic electrons.  The former is frequently used in the literature but
it has to be understood as an order of magnitude approach, as there is no
compelling physical reason why one should expect the magnetic field to be in
equipartition. The latter is made possible for a few clusters by the
observation of hard X-ray (HXR) emission that, if interpreted as the outcome
of inverse Compton scattering (ICS) of relativistic electrons on CMB photons,
implies an average ICMF strength within the emitting volume in the range of
$0.2 - 0.4~\muG$ \citep{1999ApJ...513L..21F,1999ApJ...511L..21R}. However,
both the detection of HXR as well as its interpretation in terms of ICS are
still controversial \citep{2004A&A...414L..41R,2004ApJ...602L..73F}. Other
explanations have been proposed which may allow or require stronger ICMF
\citep[see e.g.][]{Blasi.Cola:1999, Atoyan:1999ek}.

Indeed, Faraday rotation measurements (RMs) of polarized radio sources placed
within the cluster, or in the background, provide significant evidence for the
presence of stronger ICMFs, in the range $1 - 10~\muG$ in the core of non
cooling-flow clusters, and of even larger strength in cooling-flow clusters
\citep[][see also Govoni et al., in
preparation]{1991ApJ...379...80K,2001ApJ...547L.111C,Taylor:2001ji,2003A&A...412..373V}. Investigations of RMs of
elongated radio-sources within galaxy clusters also provide invaluable
information on the geometrical structure of ICMFs, which cannot be provided by
radio-halo observations alone. The data on RMs are incompatible with uniformly
oriented ICMFs, rather, a typical length-scale of $\approx 5-15$ kpc has been
inferred. Also, evidence is accumulating that there is no unique length-scale
for ICMFs, and that a successful interpretation of the RMs requires the
adoption of a power-law spectrum, even though the power law index (somewhere
in the range -1.5 to -4) is so far only very weakly constrained
\citep{2003A&A...401..835E,2003A&A...412..373V,Murgia:2004zn}.  The
discrepancy between the EGMF strengths inferred from RMs and those inferred from
the combination of radio halos and hard X-ray emission may be explained, at
least in part, with the fact that the latter is not sensitive to the presence
of magnetic substructures \citep{1999A&A...344..409E}, \citep[for a discussion about this issue and a
comprehensive review of EGMF observations see][]{Carilli.Taylor..2002}.

Another crucial issue in the present context concerns the radial
profile of ICMF in the external regions of galaxy clusters. Recent
work based on radio emission \citep{2001MNRAS.320..365B} as well as
RMs \citep{Dolag:2001} indicates that ICMFs decline in strength with
radius, with a radial profile that appears to be similar to that of
the gas density.

Outside clusters, only upper limits on the EGMF strength are
available. They are at the level of $10^{-9}-10^{-8}~{\rm G}$ for
fields extending over cosmological distances with coherence lengths in
the range of 50 to 1 Mpc, respectively \citep{Blasi.Burles..1999}. These
bounds do not hold for MFs in clustered regions, including the
filaments that connect galaxy clusters, where the field strength might
be as large as $10^{-7}~{\rm G}$. In principle, either a weak all
pervading smooth field, or stronger fields localized in a complex web
of filaments, may produce sizable deflections of UHECR over a large
portion of the sky. It is therefore evident that a better
understanding of the large-scale magnetic structure of the Universe is
called for.

\subsection{The origin of magnetic fields in galaxy clusters}\label{models}

The origin of the ICMFs is still unknown.  We can distinguish three
main classes of models that have been proposed to explain their origin.

In the first, EGMFs are assumed to be produced `locally' and at relatively low
redshift ($z \sim 2-3$) by the ejecta of galaxies
\citep[e.g.][]{Volk&Atoyan..ApJ.2000} or AGNs \citep[e.g.][]{Furlanetto&Loeb..ApJ2001}. One of
the main arguments in favor of these models is that the high metallicity
observed in the ICM suggests that a significant enrichment driven by galactic
winds or AGN must have taken place in the past.  Winds and jets should carry
MFs together with the processed matter. While it was shown that winds from
ordinary galaxies give rise to MFs which are far weaker than those observed in
galaxy clusters, ICMFs produced by the ejecta of starburst galaxies can be as
large as $0.1\,\muG$. Clearly, this class of models predicts that EGMFs are
mainly concentrated in galaxy clusters. Note that if the magnetic pollution
happens early enough (around $z \sim 3$) these fields will not only be
amplified by the adiabatic compression of the proto-cluster region, but also
by shear flows, turbulent motions, and merging events during the formation of
the galaxy clusters.

In the second class of models, the seeds of EGMFs are assumed to be produced
at higher redshift, before galaxy clusters form as gravitationally bound
systems. Although the strength of the seed field is expected to be
considerably smaller than in the previous scenario, the adiabatic compression
of the gas and the shear flows driven by the accretion of structures can give
rise to a considerable amplification of the MFs. Several mechanisms have been
proposed to explain the origin of magnetic seed fields at high redshift. Some
of these models are similar to those discussed above, differing only in the
time at which the magnetic pollution is assumed to take place. In the present
class of models the MF seeds are supposed to be expelled by an early
population of dwarf starburst galaxies or by AGNs at a high redshift between 4
and 6 \citep{Kronberg..1999ApJ}, allowing them to magnetize a large fraction
of the volume. Other models invoke processes which took place in the early
universe \citep[see][for a review]{Grasso..PhysRep.2000}.  Indeed, the
ubiquity of MFs in the universe suggests that they may have a cosmological
origin. In general, all `high-$z$ models' predict MF seeds that fill the
entire volume of the Universe. However, the assumed coherence length of the
field depends crucially on the details of the models. While models based on
phase transitions give rise to coherence lengths which are so small that the
corresponding fields have probably been dissipated, MFs generated at neutrino
or photon decoupling have much higher chances to survive until the present
time. Another (speculative) possibility is that the seed field was produced
during inflation. 
In this case, the coherence length can be as large as the
Hubble radius.

The third scenario is that the seeds of ICMFs were produced by the
so-called Biermann battery \citep{Kulsrud..ApJ.1997,Ryu..1998}
effect. The idea here is that merger shocks produced by the
hierarchical structure formation process give rise to small
thermionic electric currents which, in turn, may generate magnetic
fields. The battery process has the attractive feature to be
independent of unknown physics at high redshift. Its drawback is that,
due to the large conductivity of the IGM, it can give rise to at most
very tiny MFs, of order $10^{-21}$ G. One therefore needs to invoke a
subsequent turbulent dynamo to boost the field strength to the
observed level. Such turbulent amplification, however, cannot be
simulated numerically yet, making it quite difficult to predict how it
would proceed in a realistic environment.  It is clear that one
expects the level of turbulence to be strongly dependent on
environment, and that it should mostly appear in high density regions
like collapsed objects.  While energetic events such as mergers of
galaxy clusters can be easily imagined to drive the required levels of
turbulence, this is harder to understand in relatively quiet regions
like filaments.  Lacking a theoretical understanding of the turbulent
amplification, it is therefore not straightforward to relate the very
week seed fields produced by the battery process with the magnetic
fields observed today. Attempts to construct such models based on
combining numerical and analytical computations 
were not successful so far in reproducing the
observed scaling relations of magnetic fields in galaxy clusters.

\subsection{Simulations of magnetic field evolution in galaxy clusters
\label{SecPrevSims}}

As discussed above, a magnetic seed field generated at high redshift is
amplified during the formation of galaxy clusters due to the complex dynamics
of the gas in which the MF is frozen in.  This process has been simulated in
previous work \citep{1999A&A...348..351D,Dolag:2002} by combining
MHD with smoothed particle hydrodynamics (the `MSPH'
technique), and with the self-gravity of the gas and an additional
collisionless dark matter fluid.  It has been shown that MF amplification
takes place both due to adiabatic compression and magnetic induction, with the
latter being driven by shear flows that are ultimately powered by anisotropic
accretion and merger events.

The fact that the anisotropy of the collapse of galaxy clusters gives
rise to additional amplification of magnetic fields has also been
demonstrated with analytic models \citep{2003MNRAS.338..785B}.  Other
simulation work confirmed that merging events of galaxy clusters are a
strong driver of magnetic field amplification
\citep{1999ApJ...518..594R}, and that shear flows play an important
role in driving magnetic induction in the ICM
\citep{1999ApJ...518..177B}.  When applied to the formation of galaxy
clusters, these processes produce a magnetic field with a radially
varying strength.  In the outer region of a cluster, the field is
approximately proportional to the density of the gas, while in its
core the field profile flattens 
compared to the density.

One of the main achievements of the MSPH simulations lies in their ability to
predict not only the radial magnetic profiles of clusters but also the
geometrical structure of the field, which allows the production of synthetic
RM maps which can be readily compared with observational data. In doing this,
the simulations succeeded in reproducing, or predicting, a number of
observational facts. First of all they predicted a nearly linear correlation
between X-ray properties and RMs, which was later nicely confirmed by
observations \citep{Dolag:2001}. 
This result is of particular relevance here because it shows that the
MFs observed in clusters of different masses 
are well reproduced, and therefore
may have plausibly originated in the same cosmological seed field.  Applying an
additional model for the relativistic electron population within clusters, it
was also shown that the basic properties of radio halos, including the very
steep observed correlation between cluster temperature and radio halo power,
are well reproduced \citep{2000A&A...362..151D}. Finally, the predicted radial
profiles of RMs fit the observed ones well.

The strength of the uniform seed field required to reach this agreement was
$(1-5) \times 10^{-9}\,{\rm G}$ at redshift $z_* \simeq 20$. In unclustered
parts of the intergalactic medium (IGM), this seed field corresponds to $B_0
\equiv B(z_{*}) (1 + z_{*})^{-2} \simeq (0.2-1) \times 10^{-11}$ G at the
present time.  This range of values is well within the reach of several of the
models for magnetic seed field generation that we discussed in the previous
section.

It is important to note that due to the chaotic nature of the cluster
accretion process, no memory of the initial field configuration was
found in the synthetic ICMFs down to the scales resolved by the
simulations. This implies that the results of the simulation depend
effectively only on one free parameter, namely the comoving strength
of the seed field $B_0$. Also, since most of the magnetic field
amplification takes place at low redshift ($z \simleq 3$), the result
of the simulation does not depend on the precise time at which the
seed field is switched-on, provided it is generated before the first
significant major merger events.
Within our simulations we found, that at $z=3$ the magnetic seed
field gets amplified by at most 30\% above the expected adiabatic 
value by the anisotropic collapse within the proto-cluster regions.
Therefore, within this scenario it
is possible that the seeds of ICMFs were either existing already well
before cluster formation, or were generated later, in a local fashion,
at $z \sim 3 - 5$. 

\section{Constrained MSPH simulation of the Local Universe}

\subsection{Initial condition for the magnetic field}

As we summarized in Section~\ref{SecPrevSims}, the results of MSPH simulations
suggest that the MFs observed in galaxy clusters are consistent with
originating from a single space-filling cosmological seed field, or from
locally generated seeds which have everywhere approximatively the same
strength.  Furthermore, we know that the simulated ICMFs are largely
independent of the geometrical structure of the seed field. These facts
motivate us to use a homogeneous seed field for our initial conditions.

Following \citet{1999A&A...348..351D,Dolag:2002}, we assume the
existence of a cosmological magnetic seed field generated at high
redshift. We take the field to be uniform on the scale of the
simulation, noting that this will maximize deflections of electrically
charged UHECRs.  While ICMFs are independent of the adopted seed field
orientation, we note that this will not always be true for low density
environments on larger scales where some memory of the initial field
geometry may survive.  Therefore some attention will be required when
discussing the generality of our final results.

We performed two simulations with different strengths and orientations of the
MF seed. In the first simulation \citep[see][]{Dolag:2003ra} we used $B_0 =
1.0 \times10^{-11}$ G, based on earlier work that showed that for this value
the magnetic field in galaxy clusters reaches the maximum field strength still
allowed by the observations. For the second simulation, we used $B_{0} = 2.0
\times 10^{-12}\,{\rm G}$, which tries to optimize the agreement with the
observed RMs rather than maximizing the field strength. The orientation of the
seed fields was chosen perpendicular to each other in the two
simulations. This allows us to test a possible dependence of our results on
the arbitrarily chosen orientation of the seed field.

Note that our two cosmological simulations did not include additional
contributions to the magnetic field that are potentially injected into the ICM
by galactic winds or AGNs during the late stages of cluster formation. If
there is a significant contribution to the cluster magnetic field by such
processes, the overall initial cosmological seed field would have be to
reduced in order to not exceed the observational bounds.  As a further
consequence, the magnetic field in low density environments like filaments or
voids would then be smaller than predicted by our scenario. Therefore, our
predictions for the magnetic field in filaments should be taken as an upper
limit.

\subsection{Initial conditions for the density fluctuations}

\begin{table}[t]
  \begin{center}
  \begin{tabular}{lccc}
\hline 
  Cluster   & $T_{\rm obs} {\rm [kev]}$ & $T_{\rm sim} {\rm [kev]} $  \\
\hline
  Coma      & 8.2 & 6.5 \\
  Virgo     & 2.3 & 3.8 \\
  Centaurus & 3.5 & 4.1 \\
  Hydra     & 3.7 & 3.4 \\
  Perseus   & 5.6 & 6.2 \\
  A3627     & 7.0 & 4.0 \\
\hline
  \end{tabular}
  \caption{Comparison of the observed cluster temperature with the
  mass weighted temperature within $0.1\times R_{\rm vir}$ calculated
  from the simulations.}
  \label{tab:tab0}
  \end{center}
\end{table}

Our simulations use similar initial conditions as \citet{Mathis:2002}
in their study of structure formation in the Local Universe.  Their
initial density fluctuations were constructed from the IRAS 1.2-Jy
galaxy survey by first smoothing the observed galaxy density field on
a scale of 7 Mpc, evolving it linearly back in time, and then using it
as a Gaussian constraint \citep{Hoffman1991} for an otherwise random
realization of the $\Lambda$CDM cosmology.  The volume that is
constrained by the observations covers a sphere of radius $\sim 115$
Mpc, centered on the Milky Way. This region is sampled with high
resolution dark matter particles and is embedded in a periodic box of
$\sim 343$ Mpc on a side. The region outside the constrained volume is
filled with low resolution dark matter particles, allowing a good
coverage of long range gravitational tidal forces.

\citet{Mathis:2002} demonstrated that the evolved state of these initial
conditions provides a good match to the large-scale structure observed in the
Local Universe.  Using semi-analytic models of galaxy formation crafted
on top of dark matter merging history trees measured from the simulation, they
showed that density and velocity maps obtained from synthetic mock galaxy
catalogues compared well with their observed counterparts. Also, many of the
most prominent clusters observed locally can be identified directly with halos
in the simulation, and their positions and masses agree well with their
simulated counterparts.

For the work presented here, we extended the initial conditions by splitting
the original high resolution dark matter particles into gas and dark matter
particles with masses of $0.69 \times 10^9\; {\rm M}_\odot$ and $4.4 \times
10^9\; {\rm M}_\odot$ respectively. The most massive clusters in our
simulations are hence resolved by nearly one million particles. The
gravitational force resolution (i.e. the comoving softening length) of the
simulations was set to be $10\,{\rm kpc}$, which is comparable to the
inter-particle separation reached by the SPH particles in the dense centers of
our simulated galaxy clusters.

\begin{figure*}[t]
\includegraphics[width=0.99\textwidth]{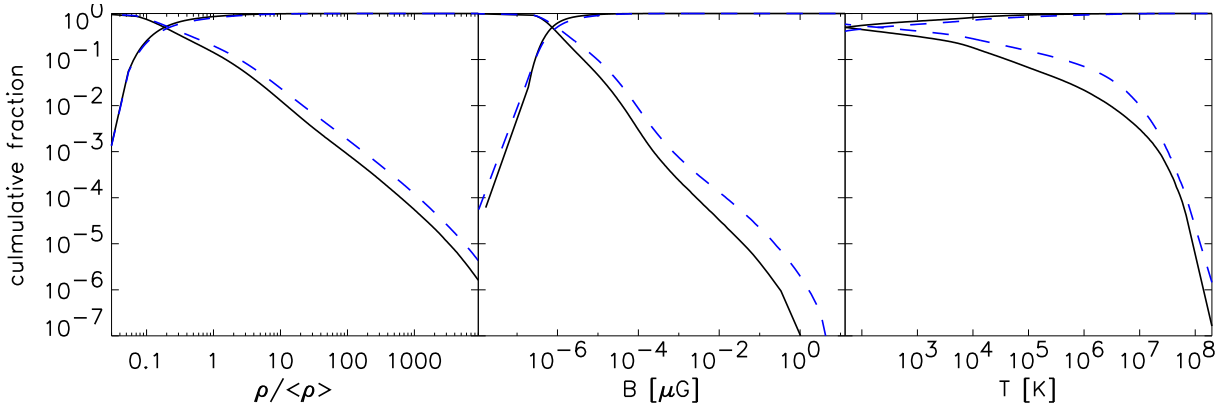}
\caption{ Volume-weighted cumulative filling factors of mean baryonic density
(left panel), magnetic field strength (middle panel) and gas temperature
(right panel). Filling factors are calculated above (decreasing curves) and
below (increasing curves) a given threshold, and are shown as a functions of
the corresponding thresholds.  We present filling factors calculated for two
boxes of 70 Mpc on a side, centered on different points in the simulation.
The solid lines are for a box centered on the `Milky Way' (observer position),
while the dashed lines are for a box centered on a void within the Centaurus
supercluster.  } \label{fig:fig1}
\end{figure*}

\subsection{Numerical approach}

Our simulations were carried out with {\small GADGET-2}, a new version of the
parallel TreeSPH simulation code {\small GADGET} \citep{SP01.1}. The code uses
an entropy-conserving formulation of SPH \citep{2002MNRAS.333..649S}, and was
supplemented with a treatment of magnetic field in the framework of ideal MHD,
as described in \citet{1999A&A...348..351D}.  Besides following the induction
equation for the magnetic field, we take magnetic back reaction into account
using a symmetric formulation of the Lorentz force based on the Maxwell
tensor. The treatment of magnetic fields in SPH was improved by explicitly
subtracting the part of the magnetic force which is proportional to the
divergence of the magnetic field, as described in
\citet{2001ApJ...561...82B}. This helps to keep the level of any numerically
induced divergence of the magnetic field at negligible values, and it also
helps to avoid a well-known instability of our MHD formulation in regions
where magnetic field pressure substantially exceeds the thermal pressure.

Note that in simulations without radiative cooling, like the ones we carry out
here, the magnetic field pressure stays well below the thermal one, even
within the cores of the most massive galaxy clusters. In very strong shocks,
it can however still happen that the magnetic field is compressed so
substantially that magnetic forces dominate the thermal ones for brief periods
of time. Such situations are handled more accurately with our new formulation.
In particular, test simulations with magnetic shock tubes showed that the use
of the entropy-conserving formulation of SPH noticeably reduces the noise level
in the solution of the MHD equations.

\section{Results}

\subsection{General properties of the simulation}

As in \citet{Mathis:2002}, we find that dark matter halos identified in our
simulations match real observed galaxy clusters in the Local Universe well in
their masses and positions. Since our simulations now also include the cosmic
gas, we can extend this comparison to properties of the intra-cluster medium
(ICM).  In Table \ref{tab:tab0}, we compare the ICM temperature of local
clusters (derived from X-ray observations) with the temperatures of the
corresponding objects in our hydro simulation.  While the agreement is not
perfect, it can still be considered to be quite good, given in particular that
the details of the hydrodynamics involved in the cluster formation process are
not directly restricted by the constraints imposed by the IRAS 1.2-Jy
survey. Reassuringly, the comparison hence suggests that the large-scale
structure in the hydrodynamic component of our simulation can serve as a good
model for the real structure within the Local Universe.

\begin{figure*}
\includegraphics[width=0.33\textwidth]{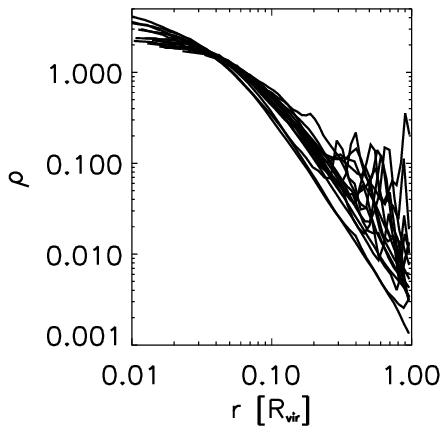}
\includegraphics[width=0.33\textwidth]{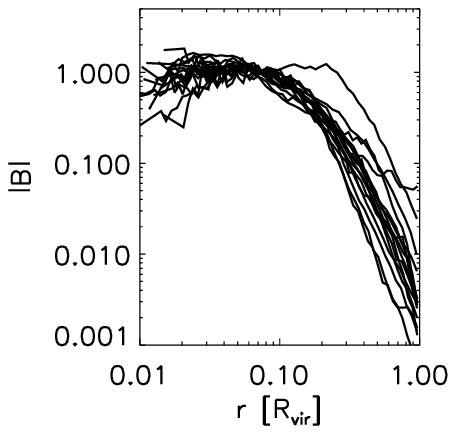}
\includegraphics[width=0.33\textwidth]{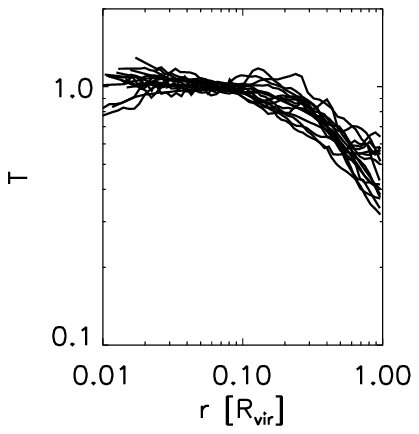}
\caption{Spherically averaged profiles of gas density (left panel),
  magnetic field strength (middle panel) and temperature (right panel,
  mass-weighted) for the 16 most massive clusters extracted from the
  simulation. The profiles are scaled to $R_{\rm vir}$ and normalized
  to have the same mean value within $0.1\times R_{\rm vir}$. }
\label{fig:fig3}
\end{figure*}

In Figure~\ref{fig:fig1}, we show the cumulative volume-weighted
distribution functions of baryonic density, magnetic field strength,
and gas temperature for a (70 Mpc)$^3$ box extracted around the `Milky
Way' (our observer position). Except for the cluster cores (highest
densities), it is evident that the distributions of density and
magnetic field are very similar, even at low densities outside of
galaxy clusters.  For comparison, we also show measurements for an
equally sized box, but centered on a relative void within the
Centaurus super-cluster. In this environment, the filling factors
inferred for the same threshold in density (or magnetic
field strength) are roughly one order of magnitude larger. Note that
for both positions, the nearest galaxy clusters are located at a
distance of $\sim 20$ Mpc. This demonstrates the importance of
choosing a realistic observer position within a comparable environment
if one wants to quantitatively assess the effects of the EGMF around
the Milky Way.

\subsection{Predicted MF scalings in galaxy clusters}\label{results1}

As we discussed in Section \ref{obs}, galaxy clusters are the only structures
where EGMFs are actually observed. To verify the consistency of our simulated
model it is therefore important to investigate its predictions for cluster
magnetic fields and to contrast them with observational data.

We begin by considering radial profiles of the gas density,
temperature, and magnetic field in galaxy clusters formed in our
simulation. In Figure~\ref{fig:fig3}, we show the mass-weighted
averages of these quantities computed for gas particles contained in
spherical shells centered on the halo centroids. The median 
density profile is compatible with the canonical $\beta$-model, 
\be 
n_{\rm gas}(r) = n(r_c)
\left( 1 - \frac{r^2}{r^2_c}\right) ^{-3\beta/2}~, 
\label{eqn:beta}
\ee 
for $ r > r_c \simeq 250~\kpc$. Consistent with earlier results
\citep{Dolag:2002}, we find that the magnetic field profiles follow, on
average, the density profiles of clusters in their outer parts, whereas in the
central regions the magnetic field profiles flatten. This flattening is
presumably a direct consequence of the lower gas velocities in the cluster
cores, which make magnetic induction less effective.

Depending on the cluster and its dynamical state, the slope of the
magnetic profile in the outer parts scatters somewhat around the slope
of the gas density. Also, the size of the magnetic core shows
substantial system to system variation. On average, however, $B$ is to
first order proportional to $n_{\rm gas}$. As mentioned in
Section~\ref{obs}, such a scaling can be tested by comparing predicted
and observed correlations between the X-ray surface brightness and the
RMs.

The amplification of the magnetic field within our simulations is not
only due to the adiabatic compression of the gas but also due to
magnetic induction driven by shear flows. If induction is important,
we expect that the final magnetic field in galaxy clusters will
depend on their merging history.  In particular, more massive
clusters, which undergo more numerous and more energetic merger
events, should end up with a higher final magnetic field in their
cores than less massive ones. On the other hand, pure adiabatic
compression should not lead to a significant dependence of the central
magnetic field on cluster mass or cluster temperature, because the
central gas density within galaxy clusters does not depend strongly on
the mass of the system.

\begin{figure}[t]
\includegraphics[width=0.45\textwidth]{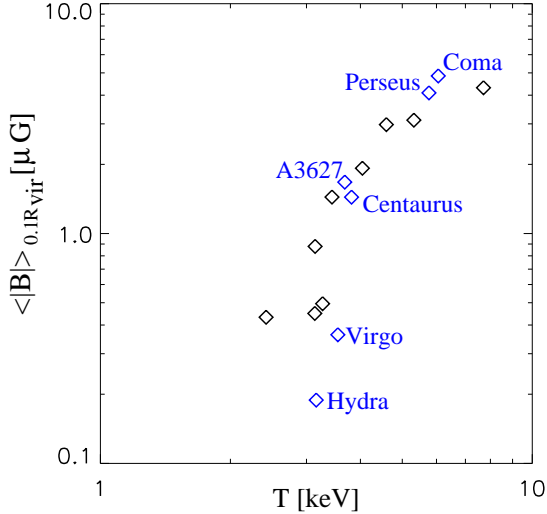}
\vspace{-0.3cm}
\caption{Magnetic field strength vs.  temperature for the most massive
clusters of the simulation.  We have labeled those halos that are
identified with actual clusters. The magnetic field has been
mass-averaged within $0.1\times R_{\rm vir}$, resulting in mean values
representative for the cluster cores. } \label{fig:fig4}
\end{figure}

Figure \ref{fig:fig4} shows the mass-averaged magnetic field calculated within
$0.1\times R_{\rm vir}$ as a function of the gas temperature in the cores of
the most massive clusters found in the simulation. It is evident from this
figure that the simulation predicts a strong dependence of $\langle B \rangle$
on temperature, showing that the build-up of the magnetic field is dominated
by induction and not compression.  Note that the slope of the $\langle B
\rangle$-$T$ correlation predicted by our simulation is comparable with that
found in earlier work \citep{Dolag:2002}, but the normalization is higher.
This is not unexpected, as in the present study the numerical mass-resolution
is 20 times better, such that the magnetic field can be tangled on much
smaller scales. In this case the value of $\left<|\vec{B}|\right>$ can be
larger, without increasing the predicted RM values.

The power spectrum of the magnetic field in the largest clusters of this
simulation has been determined by \citet{Rordorf:2004jp} finding that, on
average, it is slightly steeper than a Kolmogorov spectrum.  Significant
scatter has been found about this mean behavior, which may depend on the
cluster merger history.  We remark that the resolution limit of our
simulations is already close to (to within a factor of two), but still
slightly larger than, the smallest length scales inferred for the magnetic
field based on high resolution mapping of elongated radio sources within
galaxy clusters.

\subsection{Comparison with observations}

Our improved numerical resolution allows us to take the tangled structure of
ICMFs and its consequences for RMs much better into account than was 
possible in
previous work. It is therefore interesting to repeat and extend previous
comparisons of MSPH simulations with observational data
\citep{Dolag:2001,Dolag:2002,2000A&A...362..151D}.  For the first time we are
also able to study clusters extracted in a volume-limited fashion from a
cosmological simulation, instead of having to work with a sample of cluster
re-simulations with a poorly defined selection function.  Thanks to the
constrained nature of our simulation, we can also use the simulated `Coma'
cluster directly as a reference and calibration point. In the following, we
therefore consider in some detail a comparison with data on RMs, X-ray
correlations, and radio halos, respectively.

\subsubsection{Rotation measure profiles}

\begin{figure}[t]
\includegraphics[width=0.45\textwidth, height=0.45\textwidth]{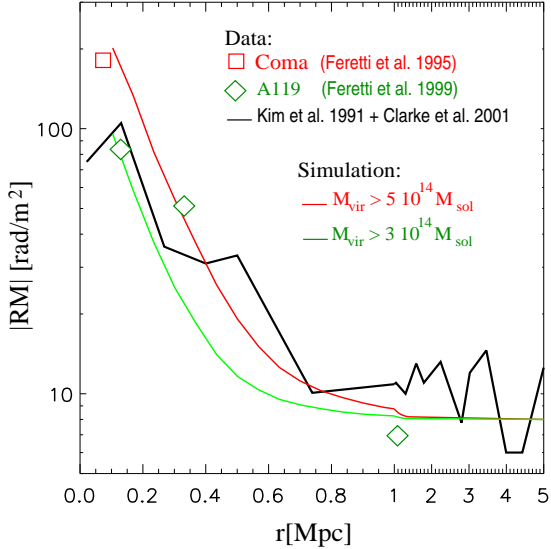}
\caption{ Comparison of RMs from the simulation with observations for Abell
clusters, as a function of distance to the closest cluster. Smooth lines
represent median values of $|{\rm RM}|$ produced by simulated clusters with
masses above $5 \times 10^{14}~{\rm M}_\odot$ and $3 \times 10^{14}~{\rm
M}_\odot$. The broken line represents the median of combined data taken from
the independent samples in \citet{1991ApJ...379...80K} and
\citet{2001ApJ...547L.111C}. We also include data (diamonds) for the three
elongated sources observed in A119 \citep{1999A&A...344..472F}, and for the
elongated source observed in the Coma cluster \citep{1995A&A...302..680F}.}
\label{fig:fig5}
\end{figure}

In Figure \ref{fig:fig5}, we compare simulated RMs with observational data
constructed from two independent samples of rotation measurements of
point-like sources in or behind of Abell clusters \citep[taken
from][]{1991ApJ...379...80K,2001ApJ...547L.111C}.  We plot the absolute value
of the RM as a function of distance from the centroid of the closest Abell
cluster. The solid line gives the median of the distributions in each of the
15 bins we used for binning the data.

Note that due to the construction of these samples, the underlying selection
function for the contributing galaxy clusters is ill-defined. This is of
importance, as the RM signal depends on cluster mass. One therefore expects a
dependence of the median profile on the mass function of the selected
clusters. As the cluster sample is composed out of Abell clusters, it is
however reasonable to assume that it mainly consists out of comparatively
massive galaxy clusters.

We also included in the plot the values inferred from three elongated sources
(diamonds) observed in the single galaxy cluster A119
\citep{1999A&A...344..472F}. This cluster compares well with the median
profile. Having a temperature of $\simeq 5.6\,$keV it also indicates that the
RM profile extracted from the observations reflects the profiles of massive
galaxy clusters. We also added one data point from the elongated source
observed in the Coma cluster \citep{1995A&A...302..680F}. Coma has a slightly
higher temperature ($\approx 8.2\,$keV) but fits well into the global trend
for massive clusters.

We calculated synthetic RM maps of the most massive clusters
performing line of sight integrations of
\be
  {\rm RM} = 812\,\frac{{\rm rad}}{{\rm m}^2}\,
  \int \frac{n_{\rm e}}{{\rm cm}^{-3}}\,
  \frac{B_\parallel}{\mu{\rm G}}\,\frac{\dd l}{{\rm kpc}}\;,
\label{RM}
\ee
where $B_\parallel$ is the magnetic field component along the line of sight
and $n_{\rm e}$ is the electron density which is found assuming complete
ionization and primordial H/He composition. 

For comparison of the observed profile with the expectations based on
synthetic RM maps obtained from our simulations we have considered two
different lower mass cuts for selecting the simulated clusters, $5\times
10^{14}\,{\rm M}_\odot$ and $3\times 10^{14}\,{\rm M}_\odot$, respectively.
Clearly, as smaller clusters are more numerous and give rise to smaller RMs,
the prediction of the simulation will depend on the adopted mass cut. Figure
\ref{fig:fig5} shows that our simulation reproduces the combined radial
profiles of the RMs quite well, as well as that for the individual cluster
A119, using a reasonable mass-cut within the range $3 - 5 \times 10^{14}~{\rm
M}_\odot$. Note that the observations do not drop to zero for large impact
parameters. Based on the averaged value on scales between 1 and 3 Mpc we added
a value of 8 rad/m$^2$ to the synthetic measurements to approximately mimic
the errors and noise of the measurements.

\subsubsection{Scaling with X-ray measurements}

Another powerful test is obtained by studying the correlation between X-ray
flux and rotation measure, at the position of the radio galaxy where the
rotation measure is observed. This amounts to comparing two
line-of-sight integrals: Eq.~(\ref{RM}) for RM with 
\begin{equation}
      \int n_{\mathrm e}^2 \; \sqrt{T} \; {\mathrm d} l \; ,
\end{equation}
the latter being proportional to the  X-ray flux.
Since the $\sqrt{T}$ dependence can be
safely neglected here, this allows us to measure how (on average) the magnetic
field scales with density. Furthermore, as we compare the same line-of-sight
for both measures, this test is independent of model assumptions for the
density distribution, or a possible asymmetry of the galaxy cluster.

We have measured this correlation from synthetic observations of our clusters,
and fitted the results obtained for individual clusters by power laws. We find
the slopes $\alpha$ to vary between 0.8 and 1.1, in agreement with
previous results \citep{Dolag:2001}.
On the other hand, from the assumptions that the gas density $n_{\mathrm gas}$
in the cluster follows a $\beta$-profile, Eq.~(\ref{eqn:beta}), and that the
magnetic field strength is scaling with the gas density as $B \propto
(n_{\mathrm gas})^\gamma$, it follows \citep{Dolag:2001} 
\be \alpha = 0.5\,
\frac{3\beta(1+\gamma)-0.5}{3\beta-0.5}\; .  
\ee 
For the usual value of $\beta=0.75$ and a magnetic field proportional to
$n_e$ (i.e. $\gamma=1$) this leads to $\alpha \approx 1.1$.

\begin{figure}[t]
\includegraphics[width=0.45\textwidth]{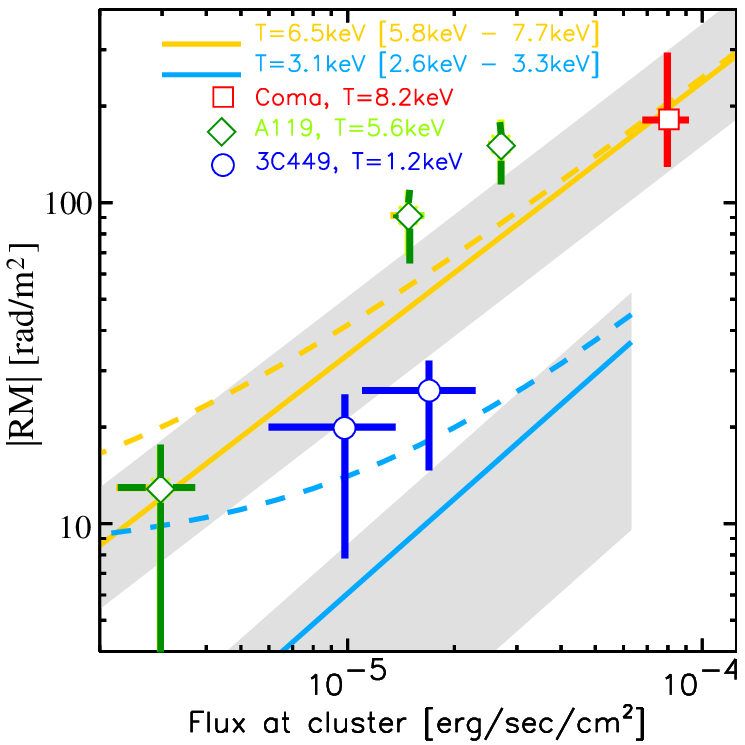}
\caption{Correlation of rotation measurements with X-ray surface brightness
for different galaxy clusters. The data points are observations for the two
massive clusters A119 \citep{1999A&A...344..472F} and Coma
\citep{1995A&A...302..680F}, and for the very poor ($T\approx 2\,$keV) system
3C449, which has a central radio source \citep{1999A&A...341...29F}. The
predictions from the simulations are obtained by averaging over the 5 most
massive clusters (upper gray region), or all clusters in the range
$2.5-3.5\,{\rm keV}$ (lower gray region). The dashed line represents the
predictions from the simulations, if we account for errors and noise within
the real observations by adding 8 rad/m$^2$.  }
\label{fig:lxrm}
\end{figure}

Figure \ref{fig:lxrm} compares the observed relation between surface
brightness and rotation measure with the results of our simulations. We
consider two sub-samples of simulated clusters, one containing the 5 most
massive ones, and the other consisting of relatively low-mass systems. The high
mass clusters span the range $6.6-8.2\,{\rm keV}$ in temperature (upper band),
whereas the lower mass systems span the range $2.8-3.5\,{\rm keV}$ (lower
band).  The region shaded in Gray marks the range of correlations we found
within these subsets, and the lines give the average correlations in the two
cases.  We included observational data points for two massive clusters, A119
\citep{1999A&A...344..472F} and Coma \citep{1995A&A...302..680F}, as well as
for a very poor system ($T\approx 2\,{\rm keV}$) hosting the central radio
source 3C449 \citep{1999A&A...341...29F}. 
Note that the data points for A119 and Coma appear somewhat differently in
Figs.~\ref{fig:fig5} and \ref{fig:lxrm}. This arises because for both
elongated sources for uniformity with other data in the first plot we
calculate the median of the RM across the source area, while in the second
plot we calculate the rms of the rotation measure.

Our magnetic field strength, which was chosen to give the best agreement with
the radial profile for RMs, lies at the lower bound of the observations in
this comparison. However, the slope of the correlation is well reproduced, and
also the trend with temperature we find in the simulations seems to be
confirmed by the observations.  More data, especially for higher cluster
temperatures, will be needed to draw definite conclusions though, particularly
since this comparison has the potential to be affected by a number of
systematic effects.  As the synthetic RMs are drawn from ideal observations,
they are free of noise and measurement errors, which could however be
significant for small observed RM values.  To illustrate this, we have added a
dashed line for the synthetic observations, representing the predictions from
the simulations if errors or noise within the real observations are roughly
modeled by adding 8 rad/m$^2$ to the synthetic observations.  Given that the
properties of the noise and the nature of systematic errors of the
observations are hard to quantify, this should be only taken as a fiducial
estimate of the potential influence of plausible measurement errors. 

Note that the X-ray observations, but also the predictions of X-ray fluxes
obtained from the simulations, are subject to systematic effects as well.  For
example, it is well known that adiabatic simulations, as used in this study,
tend to overproduce the X-ray emission of galaxy clusters compared with
simulations which also account for effects of radiative cooling and stellar
feedback. Interestingly, if we reduce the emission inferred from the
simulations by a factor of 2 to roughly account for this, a better agreement
between simulations and observations is achieved.

\begin{figure*}[t]
\includegraphics[width=0.49\textwidth]{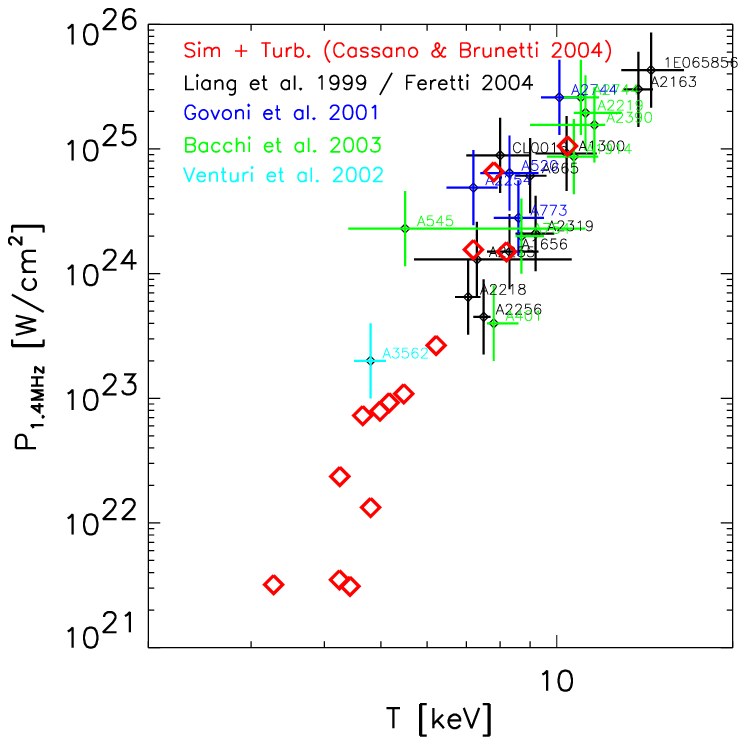}
\includegraphics[width=0.49\textwidth]{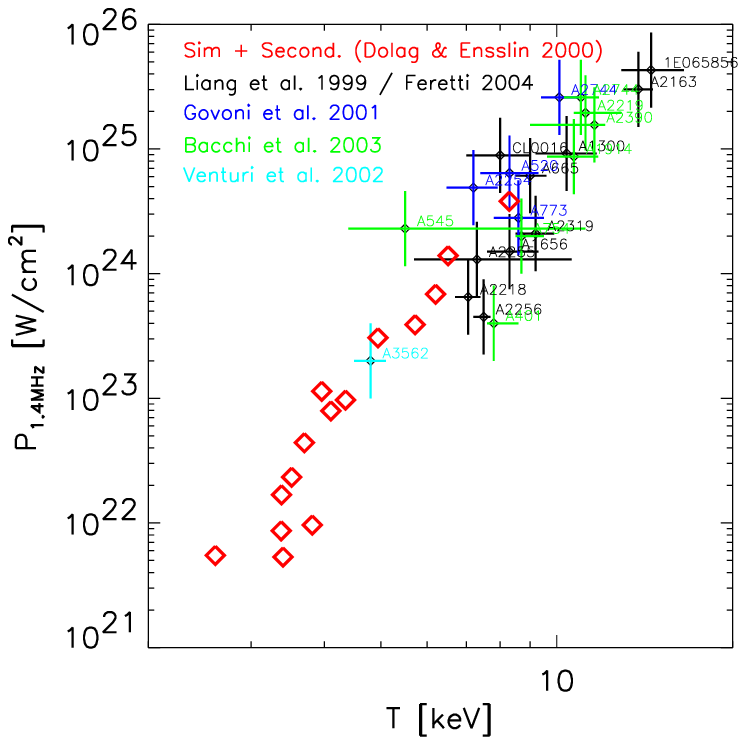}
\caption{Total power of radio halos observed at 1.4 MHz vs.
cluster temperature. We plot data from \citet{2000ApJ...544..686L},
which were partially re-observed by Feretti (2004, in preparation)
together with data from
\citet{2001A&A...369..441G,2003A&A...400..465B,2003A&A...402..913V}.
To compare with our simulations, we applied two different models for
the relativistic electron component in galaxy clusters. In the panel
on the left, we show the result obtained from a primary model based on
acceleration by turbulence (Cassano \& Brunetti 2004, in prep.), while
for the panel on the right, we used a secondary hadronic model as
described in \citet{2000A&A...362..151D}. Both models where normalized
so that our simulated Coma cluster has the same radio luminosity as
observed for the real Coma cluster. The slope of the observational
data is reproduced well by our simulated clusters in both cases.}
\label{fig:pt}
\end{figure*}

\subsubsection{Radio Halos}

The scaling relations of magnetic fields in clusters are also expected to
leave an imprint in observations of radio halos. Note that it is still
unclear how the relativistic electron component in galaxy clusters is
accelerated and maintained over the cluster volume.  The various models can
be grouped into two classes: primary models, where the relativistic
electrons are directly (re-)accelerated by various mechanism (shocks,
turbulence etc.) and secondary models, where a population of relativistic
protons is accelerated which then, by hadronic interactions with the ICM, can
produce or maintain a population of relativistic electrons \citep[for an
overview of models,
see][]{2002mpgc.book....1S,2002IAUS..199..141E,2002astro.ph..8074B}.
Independent of the model for the production of the relativistic electrons and
its dependence on the cluster temperature, the resulting emitted radio power
carries always an additional dependence on $B^2$, so that any strong relation
between magnetic field and cluster temperature will have a significant imprint
on the scaling of radio power with cluster temperature.

In recent work, Cassano \& Brunetti (2004, in preparation) developed a
semi-analytic model where they related the turbulence injected by major
cluster mergers with the acceleration of relativistic electrons. For this
model, the expected radio luminosity produced during a major merger roughly
scales as
\begin{equation}
P \propto \frac{M_{\rm vir}^3}{\sqrt{T_{\rm vir}}}
\frac{B_{\rm core}^2}{B_{\rm CMB}^2+B_{\rm core}^2}\; ,
\label{radioPower1}
\end{equation}
where we, as usual, express the electron Compton energy losses on the CMB in
terms of equivalent effective magnetic field strength, $B_{\rm CMB} = 3.24~
\mu$G.
Assuming this scaling, we can
produce a predicted correlation between radio power and cluster temperature
based on values inferred from our simulated clusters.  Figure \ref{fig:pt},
left panel, 
shows a comparison of a collection of observations with the prediction
obtained from our cluster set using the model Eq.~(\ref{radioPower1}). 
We normalized the relation
using Coma as a calibration point. For the right panel, we used a secondary
model as described in \citet{2000A&A...362..151D}, again calibrating the zero
point of the relation with the Coma cluster.  Note that in this second case,
the model is applied on a particle by particle basis when calculating the
radio emissivity by integrating over the whole cluster volume.  Both scenarios
are in good agreement with the observations, supporting the validity of our
simulations as a reasonable model for the magnetic field in real galaxy
clusters, but we cannot discriminate between different models for the origin
of the radio plasma at this point.  Interestingly, the secondary model shows
noticeably less scatter, which might be a consequence of its local definition,
which more directly accounts for the cluster geometry.

\section{The large-scale structure of cosmic magnetic fields} \label{results2}

Galaxy surveys show that clusters appear to be linked with each other
by a network of feeble {\em filaments}. These patterns in the galaxy
distribution are successfully reproduced by N-body simulations of 
CDM models, which have established the existence of a `cosmic web', 
consisting of
sheets and filaments of matter around large underdense voids.  In the
hierarchical structure formation process, dark matter and gas accrete
onto the filaments and then flow along them towards large groups and
clusters, the nodes of the cosmic web. As the gas collapses onto
filaments, we hence expect not only adiabatic compression, but also
shear flows along the filament axis, which may amplify a pre-existing
MF significantly.  In voids on the other hand, where galaxies are
almost absent and where the dark matter and gas densities are expected
to be significantly smaller than their cosmological mean values, we
expect that a pre-existing MF will only be diluted by the cosmic
expansion. No direct measurements of MFs in either filaments or voids
are available; this makes it particularly interesting to explore the
predictions of our simulation for the magnetic field configuration
within these large-scale structures.

\begin{figure}[t]
\vspace{-0.3cm}
\includegraphics[width=0.47\textwidth, height=0.45\textwidth]{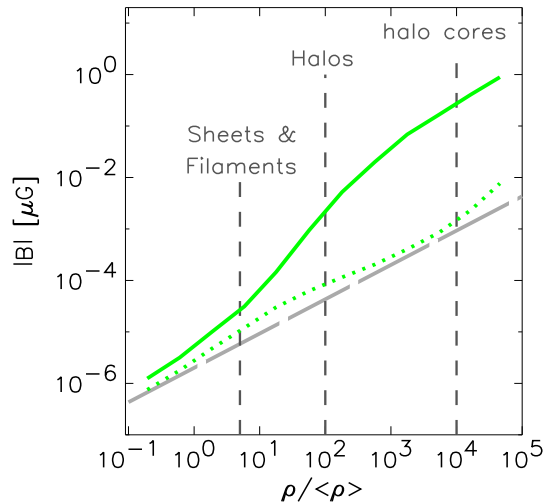}
\caption{The magnetic field strength as a function of baryonic
overdensity. The long dashed line shows the expectation for a purely adiabatic
evolution, the solid line gives the mean field strength at a given overdensity
in our simulation. While the latter is close to the adiabatic value in
underdense regions, there is a significant inductive amplification in clusters
due to shear flows and turbulence, subject however to saturation in the
cluster cores. At any given density, a large fraction of particles remains
close to the adiabatic expectation, as shown by the dotted line, which gives
the median of the distribution at each density.
\label{fig:b_ampli} }
\end{figure}

Before we consider filaments and voids in more detail, we consider in
Figure~\ref{fig:b_ampli} a global view of the field strength as a function of
baryonic overdensity. As expected, the field stays close to adiabatic
evolution in underdense regions, but already for mildly overdense structures
like sheets and filaments, a significant amplification above the value for
adiabatic compression can be observed. This inductive amplification of the
mean magnetic field becomes stronger towards the high densities reached in the
cores of clusters, where it saturates. 

\subsection{Magnetic fields in filaments and sheets}

\begin{figure*}[t]
\begin{center}
\includegraphics[width=0.32\textwidth]{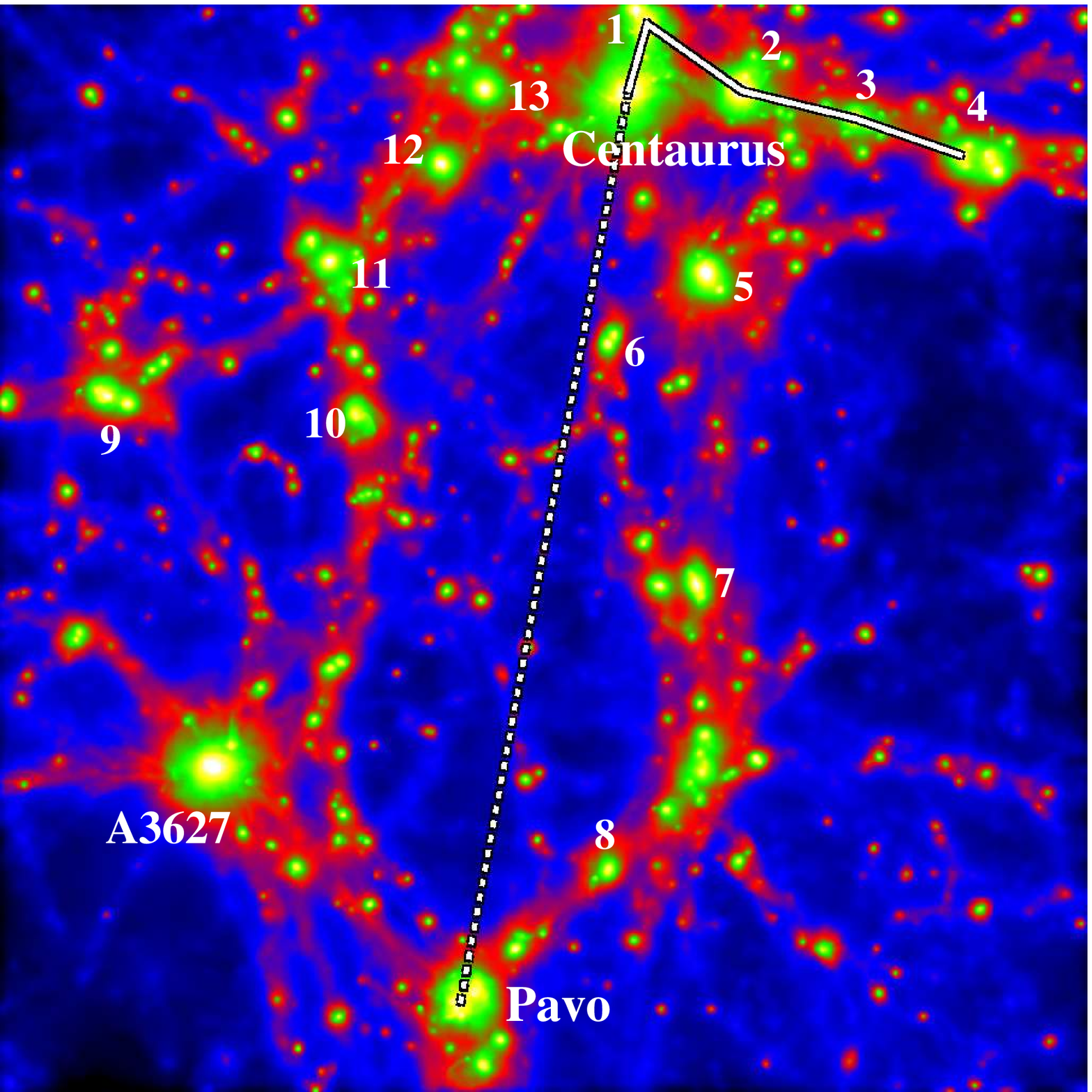}
\includegraphics[width=0.32\textwidth]{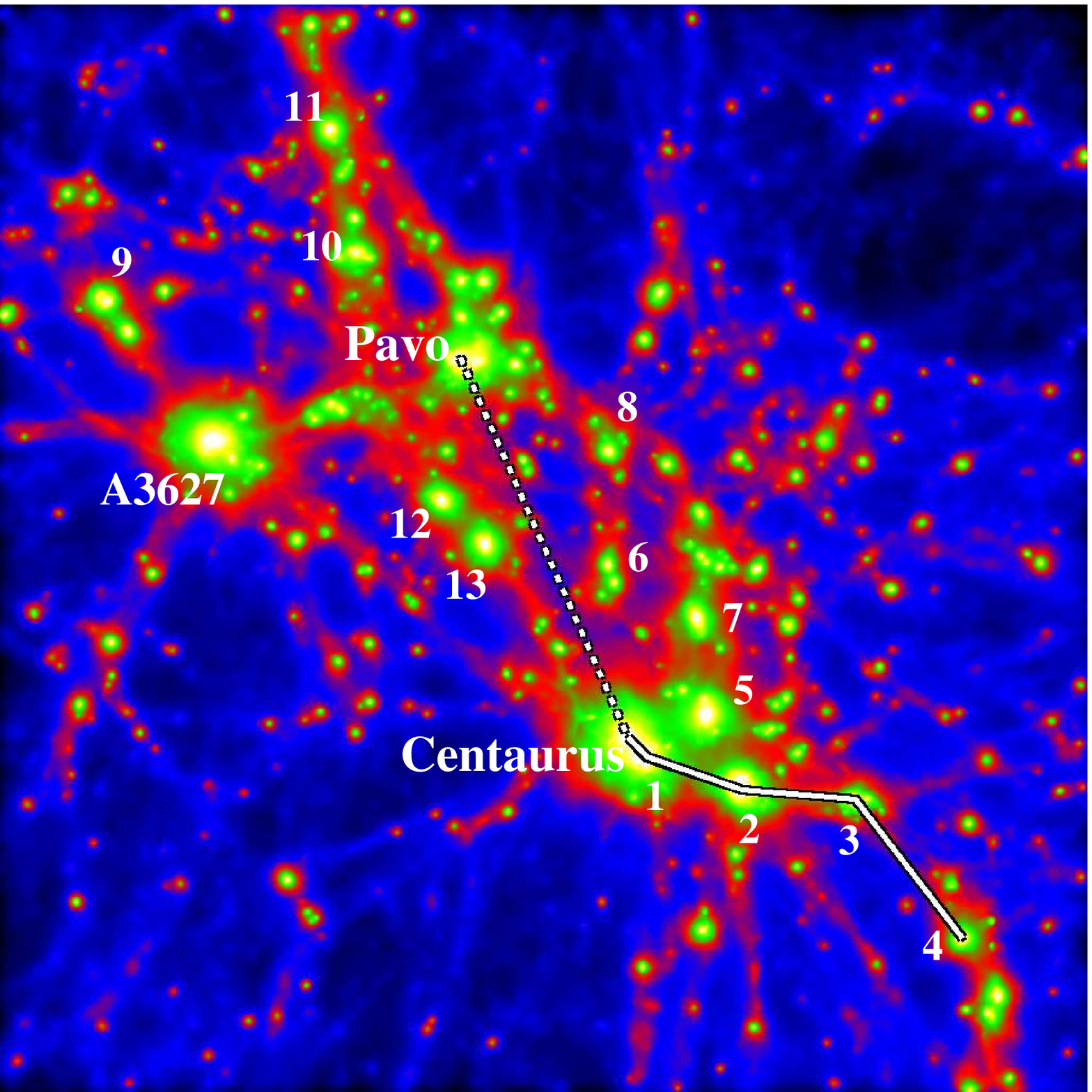}
\includegraphics[width=0.32\textwidth]{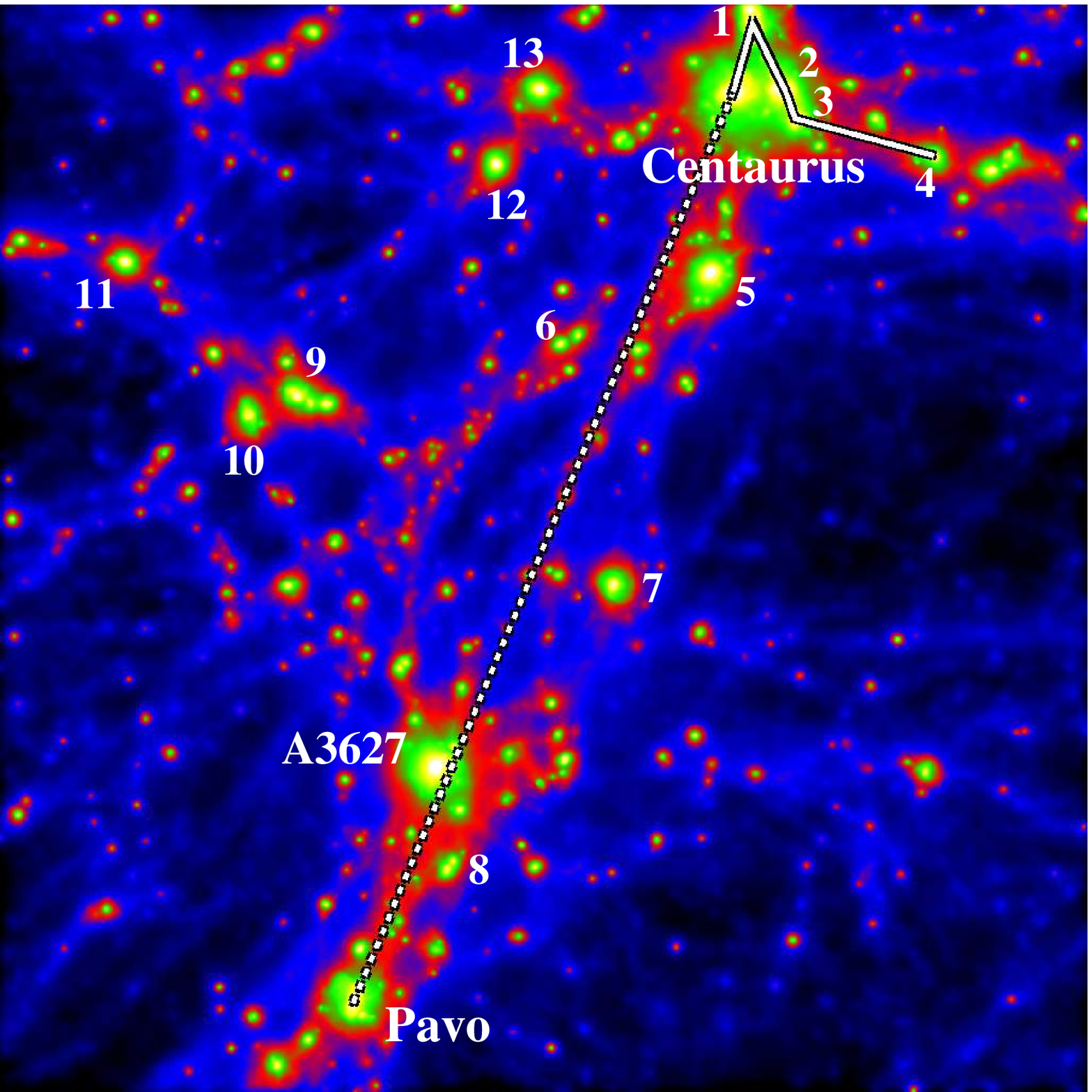}
\end{center}
\caption{Maps of the projected gas density for a region centered on a large
structure around Centaurus, Pavo and A3627. The extent of the region is 50
Mpc. Positions of several other massive halos are labeled to underline the 3D
geometry. The solid and the dotted lines mark the path along which we show
the physical state of the gas in Figures~\ref{fig:fig8} and \ref{fig:fig7}
respectively.}
\label{fig:fig6}
\end{figure*}

In the following, we focus on the Centaurus supercluster region, which gives
rise to one of the most prominent features within synthetic deflection maps
for UHECRs traversing the Local Universe (see our results in
Section~\ref{SecMaps}). 
Figure~\ref{fig:fig6} shows three projections of the gas
density of this region obtained from a (50 Mpc)$^3$ cube extracted from the
simulation.  Despite containing a huge void, which can be seen in the
left panel, the whole region still represents a large over-density. 
In fact, the average density within the region is nearly an order of
magnitude larger than the average density of an equally sized volume centered
on the position of the Milky Way (see also Fig.~\ref{fig:fig1}).

Comparatively little is known about the details of the physical state of the
baryonic gas within filaments. An important reason for this lies in their low
density, which strongly limits the resolution achieved by cosmological
simulations for the dark matter component of filaments. Note that since the
dark matter is the dominant mass component that drives via its gravity the
formation and evolution of filaments, any hydrodynamic simulation is
ultimately restricted in its resolution to that of the dark matter
component. Also, filaments are not relaxed and usually have a quite complex
geometric structure, far from being simple straight connections between
clusters.  Instead, they are often projections or crossings of sheet-like
structures that form the envelope around underdense voids. Figure
\ref{fig:fig6} demonstrates this nicely. For example, in the left panel, there
appear to be two prominent filaments which connect the {\it Pavo} cluster with
the {\it Centaurus} cluster (the left one of the two filaments includes the
cluster {\it A3627}). However, looking at different projections of the same
region (center and right panels), the situation turns out to be more complex;
what appeared as two filaments is really the projection of a complex group
consisting of several dozen halos. We have marked the 16 most massive
halos within the region to show the underlying three-dimensional structure.

\begin{figure*}
\includegraphics[width=1.0\textwidth]{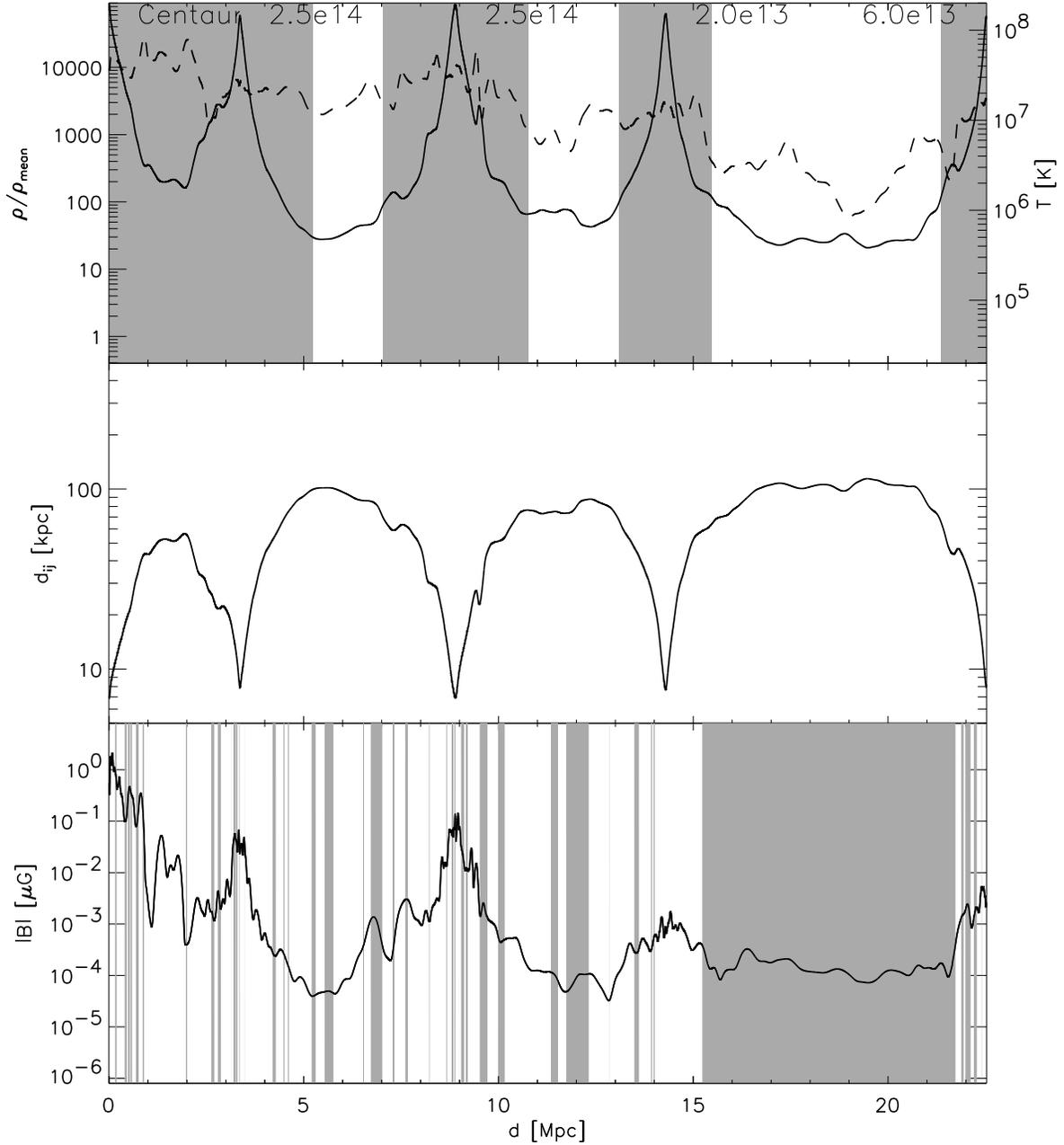}
\caption{Physical state of the baryonic matter along the filamentary
structure outlined in Figure~\ref{fig:fig6}, which starts from the
Centaurus cluster and connects to several smaller halos. The panel on
top shows the gas density as solid line, and the temperature as dashed
line. The gray region marks the virial size of the clusters. Note that
for the first two clusters, these regions overlap. The middle panel
shows the mean inter-particle separation. The panel on the bottom
gives the mean magnetic field strength. The gray stripes mark regions
where the angle between magnetic field and the line along the filament
is less than 45$^\circ$.} 
\label{fig:fig8}
\end{figure*}

In Figure \ref{fig:fig8}, we show the gas density, temperature, magnetic field
strength and orientation, along a filamentary structure connecting a chain of
5 halos which build an almost linear, 23 Mpc long structure joining into the
halo identified with the {\it Centaurus} cluster (see the solid line in
Fig.~\ref{fig:fig6} for the actual path).  This line passes through a high
concentration of clusters on a relatively short distance. Note that for the
first bridge ($\approx3.3$ Mpc long), the two virial radii overlap, so that
this part is characterized best as a high density bridge between the first two
halos. Here the lowest density within the bridge corresponds to an overdensity
$\delta$ close to 200, whereas between the last two halos, which are roughly 9
Mpc apart, the density of the filament drops to $\delta \sim 40$.  The
temperature profile along the filament is smoother and less steep compared
with the density, ranging from $10^8\,$K in the halo identified with the {\it
Centaurus} cluster to $10^6\,$K in the most tenuous regions of the filament.
Note however the numerous jumps in the temperature, in particular in the last
part of the filament, indicating the presence of several strong shocks running
along it.

To give an idea of the spatial resolution achieved within these structures, we
also plot the mean inter-particle separation of gas particles along the
filament. Given that the gas follows the dark matter well and that we have
chosen an equal number of gas and dark matter particles, this is also a good
measure of the mean inter-particle separation of the underlying dark matter
distribution. As we stressed earlier, the maximum resolution within filaments
is restricted by the mass and spatial resolution of the underlying dark matter
N-body simulation ($4.4 \times 10^9\,{\rm M}_\odot$ in our case), independent
of the spatial resolution of the hydro scheme used.

When considering the absolute value of the magnetic field along the selected
line through the 5 halos, we see that the fluctuations in field strength lie
between $10^{-2}$ and $10^{-3}~\muG$ in the high density bridge between the
first two clusters, while they go down to around $10^{-4}~\muG$ in the regions
with the lowest density contrast. We can also see that the magnetic field
strength in the low mass system is much smaller than in the high mass systems,
as expected based on the strong relation we found between mean magnetic field
in cluster centers and cluster temperature. Note in this respect that the
central density within the small mass system is close to the value found in
the high mass systems. This shows that the magnetic field amplification has
involved not only adiabatic compression but must have been driven also by
strong shear flows and merger events, which were more effective in the larger
clusters.

The regions shaded in gray mark places where the angle between the magnetic
field and the path marked in Fig.~\ref{fig:fig6}
is less than 45 degrees. Often large-scale flows twist the field but the field
strength is largely unchanged. MF rotations are most intense in
clusters, where the gas flows are in general very dynamic. At the positions of
shocks, the changes in the field direction can be very sharp, even in low
density regions.  In the high-density segment of the filament between the two
close massive clusters, the MF is chaotically oriented, suggesting that it
keeps no memory of the seed field structure.  On the other hand, in the
peripheral segment where the density stays quite low and non-linear effects
are weak, the field is coherently oriented, and reflects the structure of the
initial seed field. In this low-density environment, essentially only
adiabatic compression is operating.  Note that the $\rho^{(2/3)}$ expectation
for adiabatic amplification assumes an isotropic collapse, which is not really
true for filaments, where particularly low-density filaments maintain a memory
of the direction of the seed field.  For a given filament, we therefore expect
a dependence of the field amplification on the direction of the seed field.
However, because a cosmological simulation contains filaments of all
directions, statistical inferences like the average magnetic field strength
within filaments should be unaffected by this dependence.

\begin{figure*}[t]
\includegraphics[width=1.0\textwidth]{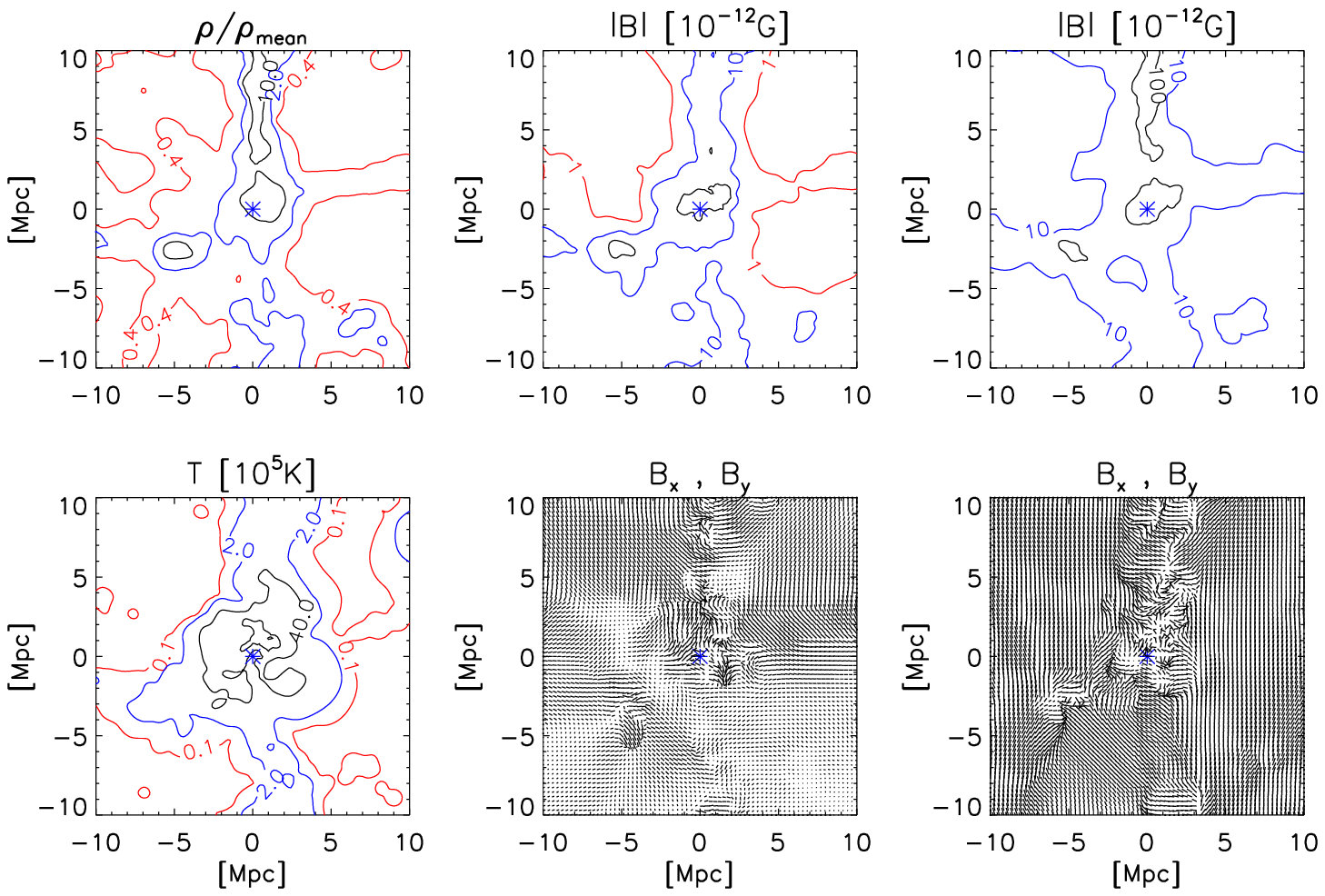}
\caption{Physical state of the gas in the plane orthogonal to a filament (the
plane is located in the middle of the last segment of the filamentary
structure shown in Figure~\ref{fig:fig8}). Two panels in the left column show
density and temperature contours, respectively. Panels in the middle column
give the strength of the magnetic field and the direction of the normalized
$B_x$, $B_y$ components for the run with the initial seed producing $B_0 =
2\times 10^{-12}$ G. The right column shows the same quantities, but for the
run with $B_0 = 10^{-11}$ G.}
\label{fig:fig9}
\end{figure*}

We now turn to examining the structure of the magnetic field orthogonal to
filaments.  To this end, we consider a plane located in the middle between the
last two halos, orthogonal to the filament studied above.  The panels in the
left column of Figure \ref{fig:fig9} show contours of density and temperature
out to a distance of 10 Mpc from the filament.  As can be seen from the
density map, the filament connects several sheets that extend from it in a
radial fashion. Interestingly, in the temperature map, only the massive sheets
are visible and the filament itself looks noticeably rounder and larger in
extent.  The middle and right panel on top show the absolute value of the
magnetic field for the two simulations we performed.  Apart from the offset in
amplitude, which is a direct consequence of the different strengths of the
initial seed fields, the structure obtained from the two simulations looks
remarkably similar and corresponds to the structure observed in the density
very well.

The two bottom panels on the right of Figure~\ref{fig:fig9} show the
orientations of the two magnetic field vector components within the plane,
normalized using all three components. The different orientations of the
magnetic field vectors in the two cases are a result of the different
orientation of the initial magnetic seed field in the two simulations. While
the distortion pattern of the field is not exactly the same in the two cases,
large similarities are nevertheless present.  The field is never really highly
chaotic in the low density filament that crosses the plane in the center, but
it is clearly more tangled in the filament core, and along the
densest sheet, which testifies to the presence of shear flows in these
structures.  In the low density parts away from the filament, the MF keeps its
coherent structure and is just slightly reoriented with respect to the initial
field.  Due to their huge extension of more than 10 Mpc and their coherent
orientation of the magnetic field, sheets and voids may then also 
produce sizeable
deflection of UHECRs, as we will discuss further in section~\ref{SecMaps}

\begin{figure*}
\includegraphics[width=1.0\textwidth]{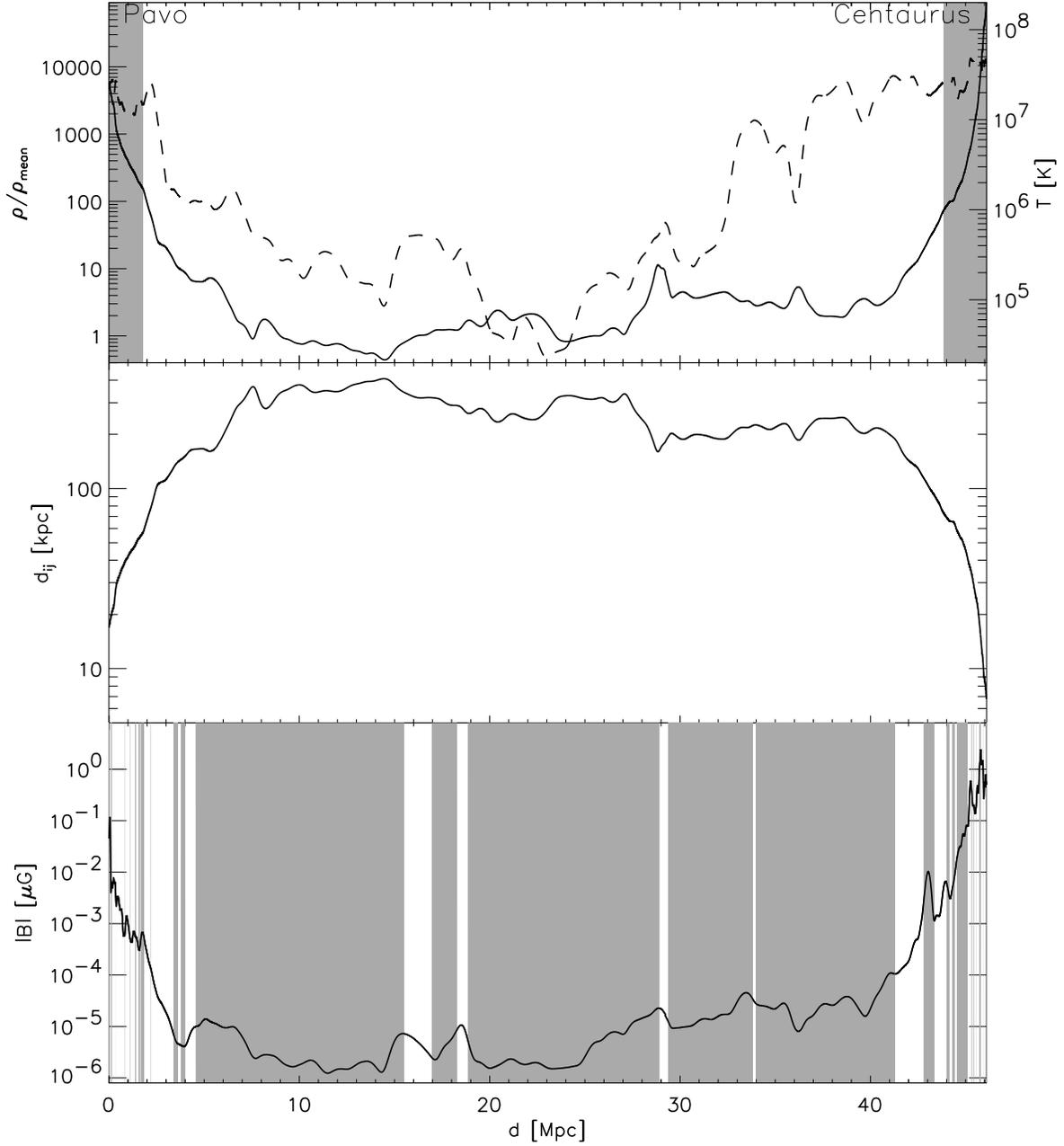}
\caption{Physical state of the gas along a straight line connecting clusters
A3627 and Centaurus. Note that this line-of-sight passes through a relative 
void
within this supercluster structure (see Fig.~\ref{fig:fig6}). The panel on top
shows the gas density as a solid line, and the temperature as a dashed
line. The gray regions marks the virial size of the clusters. The middle panel
gives the mean inter-particle separation of gas particles, illustrating the
adaptive resolution of the simulation. The bottom panel shows the mean
magnetic field strength. Here, the gray stripes mark regions where the angle
between the magnetic field and the line through the void is less than
45$^\circ$. }
\label{fig:fig7}
\end{figure*}

\subsection{Magnetic fields in low-density regions}

\begin{figure*}[t]
\includegraphics[width=1.0\textwidth]{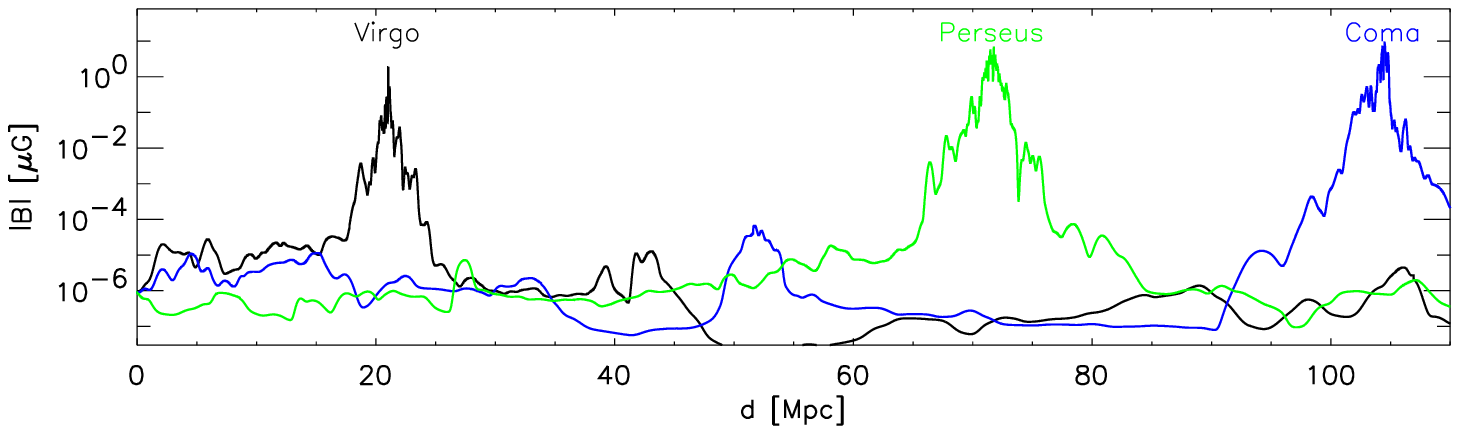}
\caption{Magnetic field strength along 3 fiducial lines through the
simulated volume. They are chosen to pass through the clusters Coma,
Virgo (which lie within the super galactic plane) and Perseus
(which is outside the super galactic plane).}
\label{fig:fig12}
\end{figure*}

To examine the structure of MFs in low-density regions, we consider the
moderate void visible within the Centaurus supercluster (see previous
section). In its longest dimension this region spans over 45 Mpc and covers the
space between the halos associated with the {\it Centaurus} cluster, {\it
A3627}, and {\it Pavo}. In order to determine the state of this vast, quite
underdense region, we measure physical quantities along a straight line
joining the halos identified as the {\it Centaurus} and {\it Pavo}
clusters. We show the results in Figure~\ref{fig:fig7}. In going away from the
clusters, the gas density decreases continuously until, at a distance of about
$10~\Mpc$, it reaches a value corresponding to $\delta$ around $1$, which it
keeps over several tens of Mpc. While also falling towards the center of the
void, the temperature shows a less well defined floor. Instead, there are
signs of heating towards the outer parts of the void. This is presumably from
hydrodynamic shock waves which have traveled into the cold low density
regions. Because the dimension of the void orthogonal to the chosen line is
smaller, such shocks may have also plausibly originated in structures `on the
side' of the void.  The MF strength in the void follows a similar profile as
the density.  Its value stays close to the comoving intensity of the seed 
field $B_0$ over most
of the void, retaining coherent orientation.
As soon as the density grows in the proximity of the clusters, the field
orientation gets randomized as a result of structure formation.

In Figure \ref{fig:fig12}, we show the strength of the magnetic field along
three additional lines-of-sight through our simulation, starting all from the
observer position at the center (i.e. the 'Milky Way'). The lines of sight
were chosen to pass through some of the prominent clusters identified within
the simulation. We selected the Coma and Virgo clusters, which lie at
different angular positions in the supergalactic plane (SGP), and Perseus,
which is along a line-of-sight almost perpendicular to this plane.  We can see
from Fig.~\ref{fig:fig12} that massive clusters affect the field over
substantial distances, but a few Mpc away from their cores, the field strength
always drops well below $10^{-4}~\muG$. Apart from a region a few Mpc wide
around Virgo, the field does not grow by more than an order of magnitude in
the SGP, and it falls even below the comoving intensity of the 
seed field $B_0$ in the direction
of Perseus. The field structure is probably highly tangled in the SGP.
Several big voids with $B \ll B_0$ are also visible in the region between the
border of the SGP and the Coma cluster. Similar as for filaments, the
adiabatic dilution of the field in a given void can depend on the relative
orientation of the magnetic seed field and the possible asymmetry in the
physical expansion of the void.  For an ensemble of voids in a cosmological
volume, this effect is however going to average out in the mean.

\section{Propagation of UHE protons in the web of extra-galactic
  magnetic fields\label{SecMaps}}

A detailed knowledge of extra-galactic magnetic fields (EGMF) is particularly
important for an understanding of the physics of UHECR. MFs influence the
propagation of UHECR primaries. The fate of both high-energy
photons\footnote{More precisely, MF are dumping the electromagnetic cascade
initiated by HE photons, strongly limiting the propagation distance and
contributing into diffuse GeV photon background, see
e.g.~\citet{Lee:1996fp,Kalashev:2001qp}. In this respect MF are important for
the propagation of UHE neutrinos as well, placing important bounds on various
models for the origin of UHECR \citep{Kalashev:2002kx}.} and charged nuclei
depends in a crucial way on the strength and spatial structure of the EGMF.

For sufficiently small propagation distances (such that fragmentation on an
infrared background is insignificant) the result for heavy nuclei can be
obtained by simply multiplying proton deflections by the nucleon
charge. However, the case of energetic photons is more complicated and
requires dedicated study.  Such an analysis is important, in particular, for a
physical interpretation of correlations with BL Lacs
\citep{Tinyakov:2001nr,Tinyakov:2001ir}, as seen in new independent HiRes
stereo datasets which apparently require a fraction of neutral particles in
UHECR \citep{Gorbunov:2004bs} at the level of a few percent.  In this work,
however, we focus on the propagation of UHE protons in the Local Universe,
leaving the case of photons and heavy nuclei for future studies.

Trajectories of charged particles in a magnetic field are bent by
the Lorentz force,
\begin{eqnarray}
\frac{\vec{{\rm d}v}}{{\rm d}t} = \frac{Ze} {E} \; \; [\vec{v} \times \vec{B}
] \; . \label{LorentzForce}
\end{eqnarray}
Using our predicted 3D map of the magnetic field within a simulation
that closely reproduces the real large-scale structure of the Local
Universe, our goal is to construct an associated map of particle
deflections under the action of this force.

The force, and correspondingly the resulting deflections, are
inversely proportional to the particle energy $E$. During the course
of propagation, all primaries are losing energy in interactions with
various cosmic backgrounds. We have calculated accumulated deflections
using two approaches. In the first approach, the energy losses are
taken into account, in the second the energy appearing in
Eqn.~(\ref{LorentzForce}) is considered to be constant. The usefulness
of the second approach will became clear in what follows.  In both
schemes, we present results as functions of energy {\it at the
detector}, not at the source. This allows a direct physical
interpretation and gives an {\it upper bound} on deflections. Indeed,
given that $E$ is an energy at detection, the original energy can only
be larger, and true integrated deflections over the whole propagation
path can only be smaller in the second scheme, where the energy losses
are neglected. In the first scheme, with unknown distances to sources,
we obtain an upper bound on deflections by placing all sources at the
maximum distance possible.

We do not follow particle trajectories directly; instead, we compute
accumulated deflections along rectilinear paths. This is a reasonable
simplification, since we are not interested in actual source
positions, but rather in finding ``windows'' where deflections are
small. We cut the deflections if they exceed 5 degrees; up to this
point the small angle approach we use is still a good
approximation. To obtain a homogeneously sampled deflection map, we
distribute sources uniformly on the sky at the maximal distance. A
``whole-sky'' map of deflections obtained in this way is simply a
particular 2D representation of the 3D EGMF structure, and should be
understood as such. With appropriate care, such maps are ready for use
in UHECR applications though.

\subsection{Deflection maps}

We considered protons with arrival energy $E = 4 \times 10^{19}~ \eV$
and $E = 1 \times 10^{20}~ \eV$. The typical distances to sources
where such primaries may originate is expected to be limited to $\sim
500$ Mpc and $\sim 50$ Mpc, respectively.

\begin{figure*}
\vspace{-0.5cm}
\includegraphics[width=0.99\textwidth]{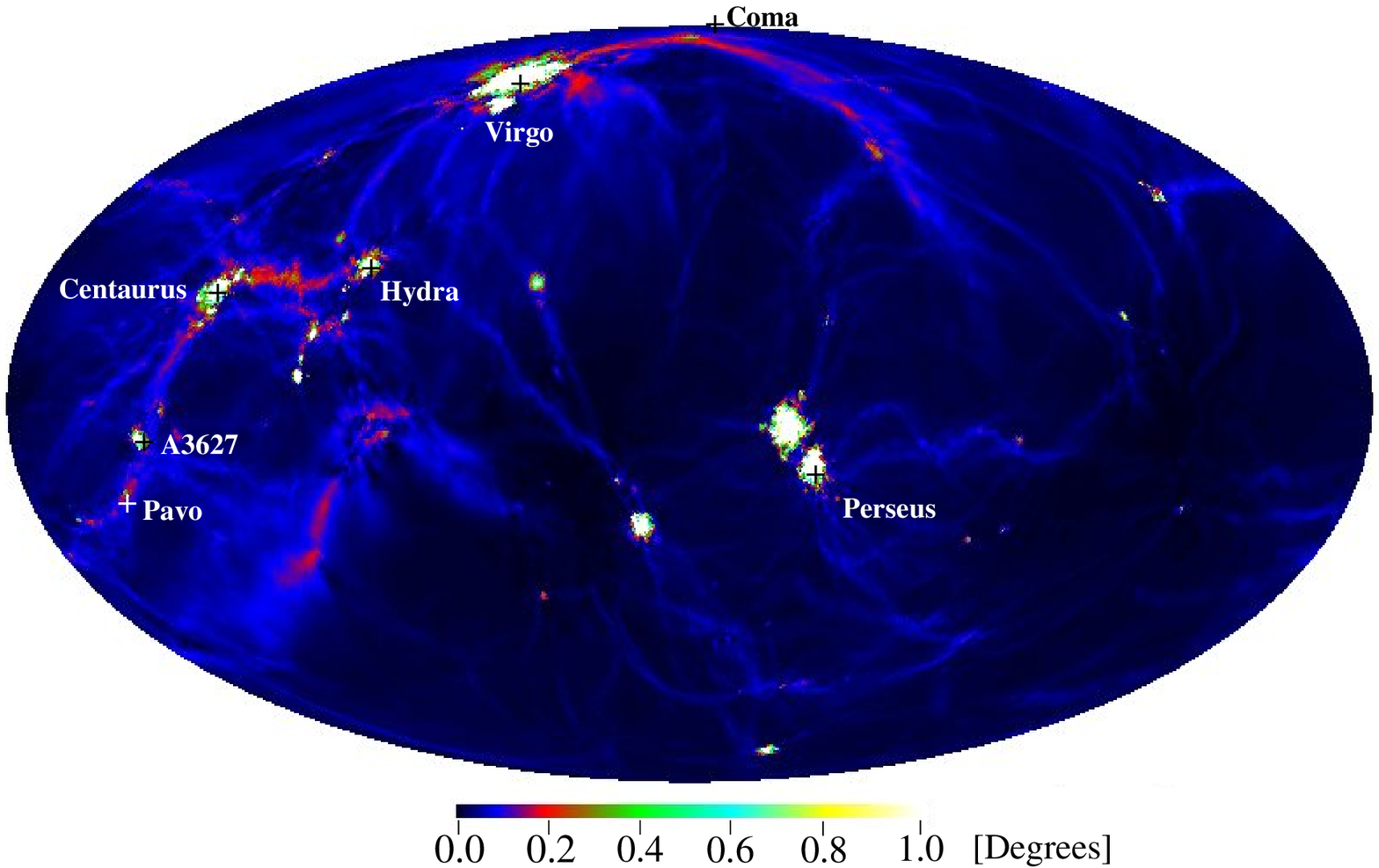}

\vspace{0.25cm}
\includegraphics[width=0.99\textwidth]{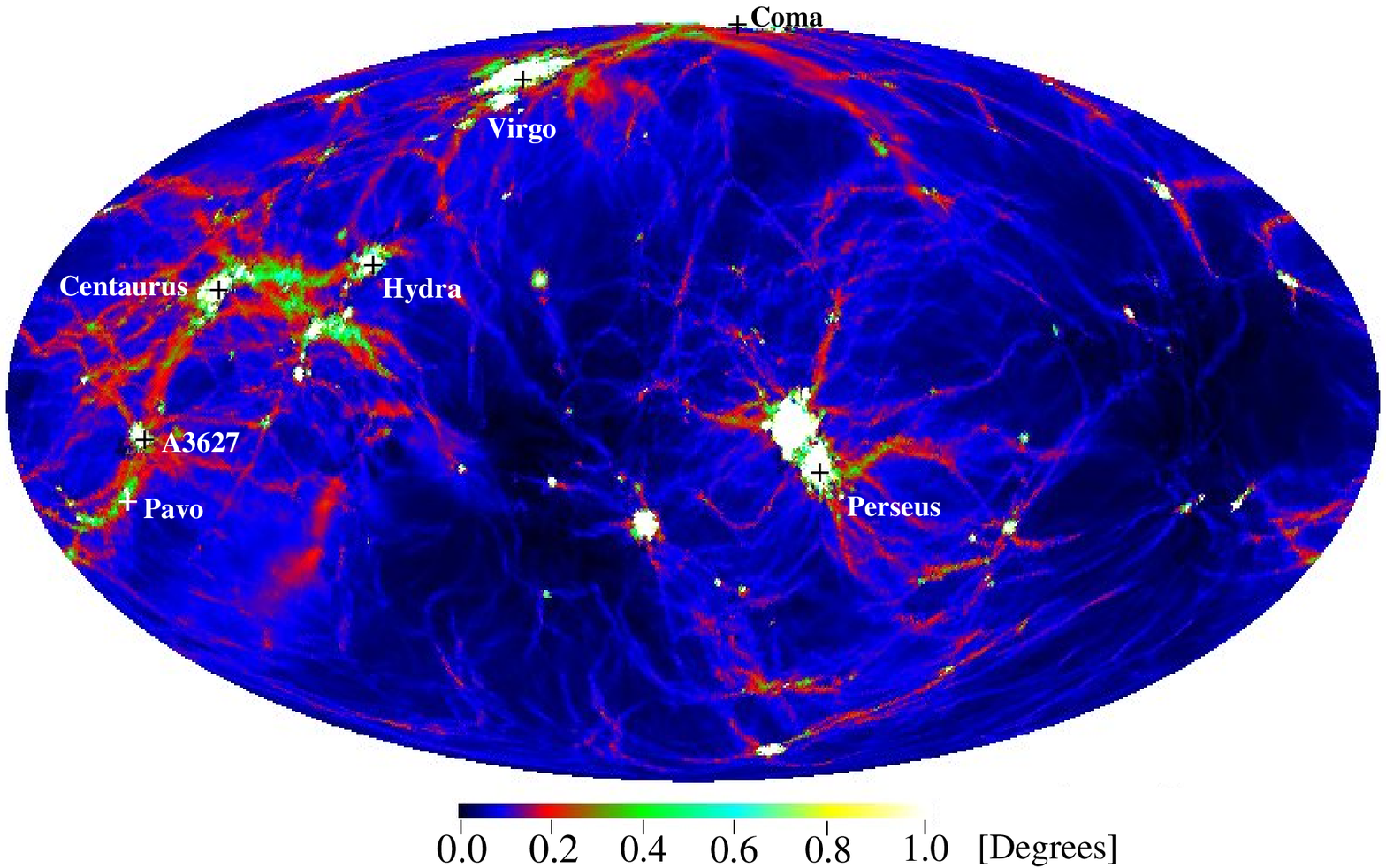}
\caption{Full sky maps of expected deflection angles for protons with
the arrival energy $E = 1\times 10^{20}$~eV. In the upper panel,
energy losses have been taken into account. The coordinate system is galactic,
with the galactic anti-center in the middle of the map.}
\label{fig:deflections1e20}
\end{figure*}

\begin{figure*}
\vspace{-0.5cm}
\includegraphics[width=.99\textwidth]{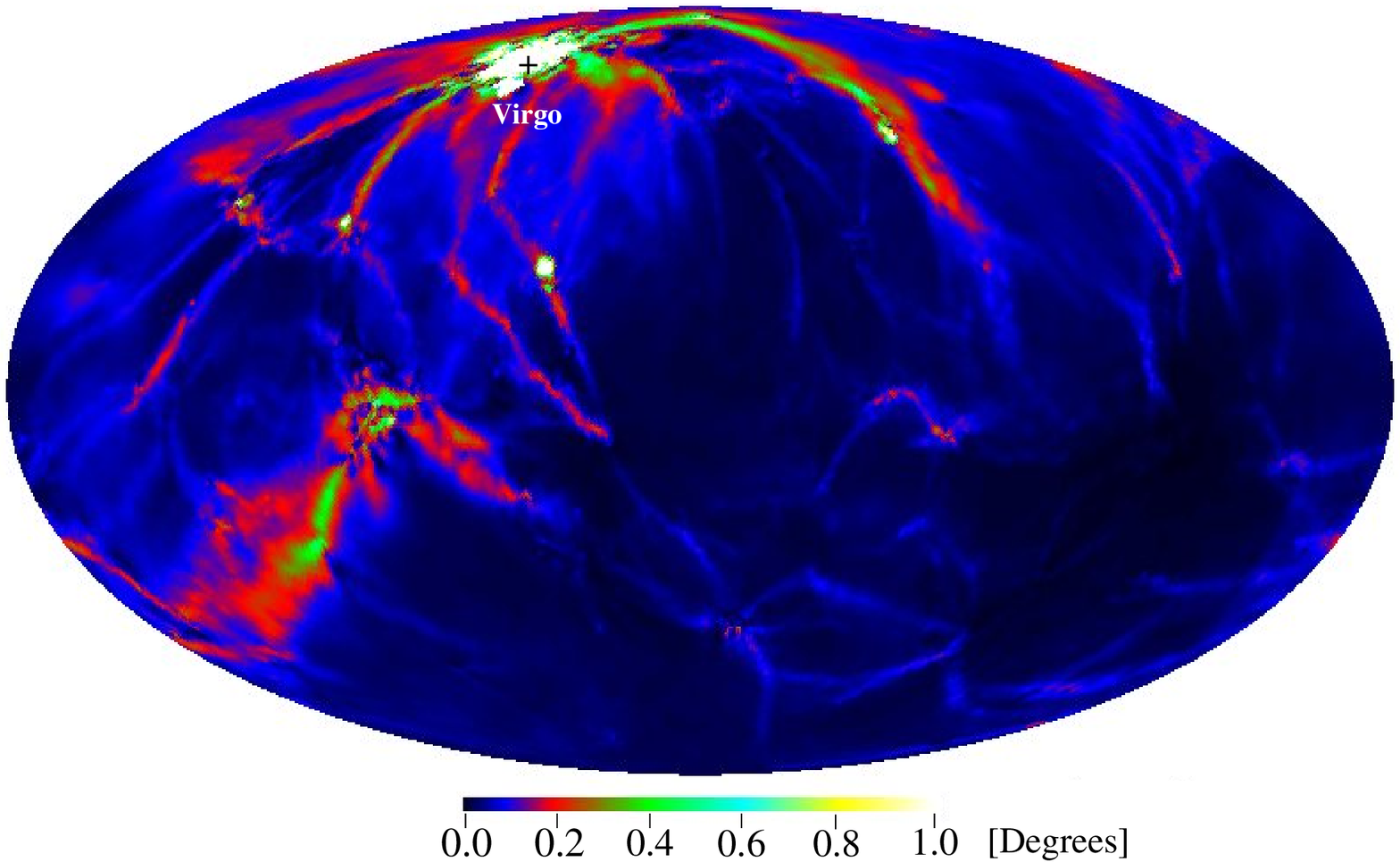}

\vspace{0.25cm}
\includegraphics[width=.99\textwidth]{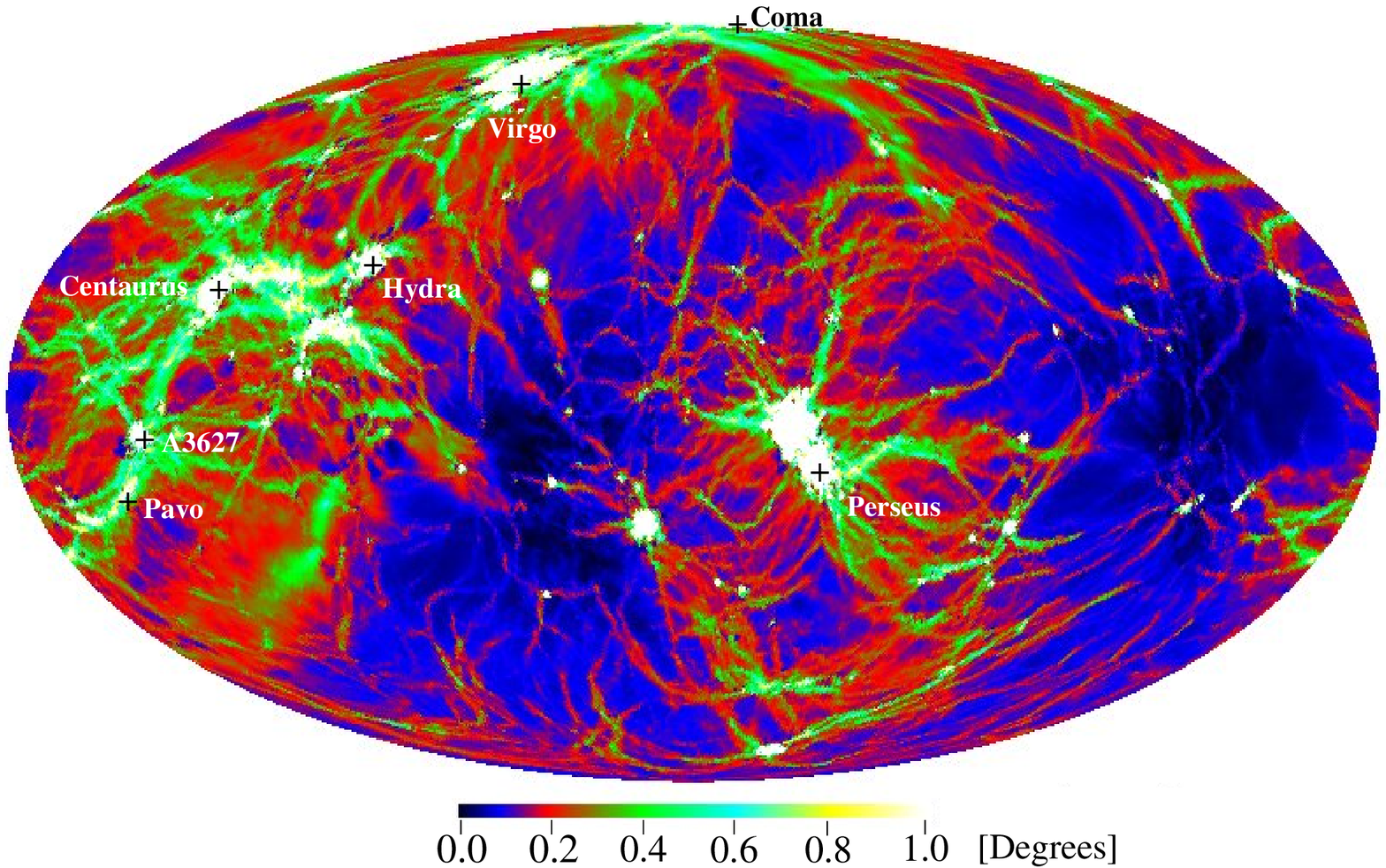}
\caption{Full sky maps of expected deflection angles for protons with the
arrival energy $E = 4\times 10^{19}$~eV. The upper panel is restricted to the
$25$~Mpc propagation distance, while in the lower panel the whole simulation
volume within a radius of $110$~Mpc around the position of the Galaxy was
used. }
\label{fig:deflections1e19}
\end{figure*}

In the case of $E = 1 \times 10^{20}~\eV$, the process of photo-pion
production in collisions with cosmic microwave background (CMB) photons, $p +
\gamma_{\rm CMB} \rightarrow p(n) + \pi^{0(+)}$, is the only important
reaction for the energy losses \citep{Greisen:1966jv,Zatsepin:1966jv}.  
We treat the corresponding attenuation of
energy analytically in the continuous loss approximation using the expressions
given in \cite{Berezinsky:1988wi} and \cite{Anchordoqui:1996ru}; for a recent
review see \cite{2002hep.ph....6072A}.  We distribute the sources over a
sphere with radius 100 Mpc around the observer location. This propagation
distance corresponds to an energy at the source of $E_{\rm source}\approx 
1 \times 10^{22}~ \eV$, 
which we argue is a generous upper limit for the maximum
acceleration energy of protons. The corresponding map of deflections is shown
in the upper panel of Figure~\ref{fig:deflections1e20}.  Smaller values of
$E_{\rm source}$ would correspond to smaller propagation distances, and hence
smaller deflections. Larger values of $E_{\rm source}$ would give rise only to
a minor change of the deflection map.

It is interesting to compare this map to the one obtained with energy losses
being neglected; the latter is shown in Fig.~\ref{fig:deflections1e20}, lower
panel. Physically, given a sufficiently large set of detected events, such a
map may give bounds on deflections in some rare cases when a particle travels
large distances without interaction, while the former map corresponds to a
statistical mean in energy losses.  In other words, assuming that the
underlying 3D model of EGMF is correct, the map presented in the lower panel
gives a conservative upper bound on deflections diminishing the caveat of
using the continuous loss approximation.

In the lower panel of Figure~\ref{fig:deflections1e19}, we show the deflection
map of protons with arrival energy $E = 4 \times 10^{19}~ \eV$, where sources
were put at a distance of 110 Mpc and energy losses were neglected.  In this
case, the choice of the maximal distance is not motivated by physical
arguments, but rather by the limited size of the simulation volume. At this
energy, the attenuation length is large, $l_E \sim 1000~\Mpc$
\citep[e.g.][]{Berezinsky:2002nc}.  Therefore, the neglect of energy losses
gives approximately a 10\% overestimate of deflections over the distance of
110 Mpc. This effect would be nearly invisible in the map of deflections. In
any case, neglecting energy losses, we obtain an upper bound for the
deflections, suitable for our purposes.

Clusters and filaments imprint a clearly visible pattern in
Figure~\ref{fig:deflections1e19}.  Large deflections are produced only when
protons cross the central regions of galaxy clusters \citep[see][for a
detailed study of this regime]{Rordorf:2004jp}, and most of these strongest
deflections are found along a strip which can be approximately identified with
the Great Attractor. The observed positions of Virgo, Coma, Hydra and
Centaurus lie in this region. Their locations quite precisely coincide with
regions where the deflections exceed $1^\circ$. Perseus and other minor
clusters produce large deflections in other well delineated regions of the
sky. Outside clusters, which occupy only a small fraction of the sky,
deflections of $0.5^\circ$ occur along an intricate network of filaments,
covering a larger area. The regions with $\delta \ll 0.5^\circ$ correspond to
voids where the MF strength is smaller than $10^{-11}$ G.

In order to investigate the relative importance of deflectors at different
distances, we also produced deflection maps that only included deflecting
magnetic fields up to some maximum distance.  We observe no significant
deflections produced at distances smaller than 7 Mpc. The deflection map
corresponding to a propagation distance of $25$ Mpc from the `Milky Way' is
shown in Fig.~\ref{fig:deflections1e19} (upper panel).  Up to this distance,
only the Virgo cluster is contributing significantly.  Massive clusters at
large distances ($\sim 100~$ Mpc) produce large deflections but cover only a
negligible fraction of the sky, so that the bulk of the deflections is
produced by passages through filaments.

We note here that if a UHECR source is placed in the core of a
galaxy cluster, the large deflections induced in the proton trajectory
while getting out of the cluster will not result in a significant
displacement, or in a smearing, of the apparent position of the
source.  This is because the angular size of clusters cores, barring a
few nearby clusters, is smaller than the angular resolution of the
UHECR experiments. For this reason, even making the extreme assumption that 
all UHECR sources are placed in galaxy clusters would have no significant 
effect on the conclusions of this paper.

An additional technical argument for neglecting energy losses is that
the flux of protons with $E = 4 \times 10^{19}~ \eV$ is expected to be
dominated by sources at distances well beyond 110 Mpc, which makes
extrapolation to larger distances necessary. Such an extrapolation can
be obtained easily by assuming a constant energy, as described in the
next subsection.

\subsection{Extrapolation to larger distances}

\begin{figure}[t]
\includegraphics[width=0.47\textwidth]{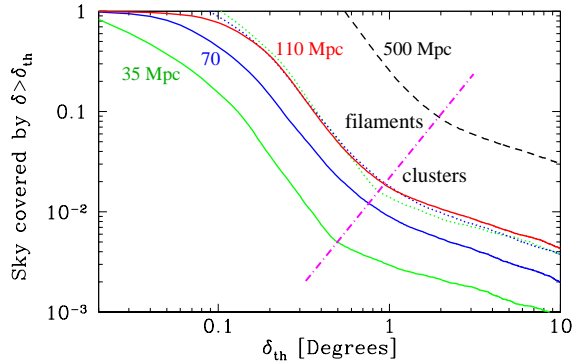}
\caption{Cumulative fraction of the sky with deflection angle larger
than $\delta_{\rm th}$, for several values of propagation distance
(solid lines). We also include an extrapolation to $500\,$ Mpc,
assuming self-similarity, based on Eqn.~(\ref{selfsimilar}) (dashed
line). To validate this scaling we show the extrapolation of
$A(\delta_{\rm th},d)$ from 35 Mpc and 70 Mpc to 110 Mpc (two dotted
lines), giving a nice agreement with the direct calculation. The
assumed UHECR energy is $4.0\times10^{19}$ eV.}
\label{histo}
\end{figure}

In Figure~\ref{histo}, we plot the fraction of the sky, $A(\delta_{\rm th})$,
over which deflections larger than $\delta_{\rm th}$ are found for different
propagation distances, as indicated in the plot. We see that deflections
larger than $1^\circ$ are to be expected over less than $2\%$ of the sky up to
a distance of $d = 110~\Mpc$.  Comparing this figure to the map of
deflections, Fig.~\ref{fig:deflections1e19}, we can easily identify those
parts of the function $A(\delta_{\rm th},d)$ which describe deflections in
clusters and filaments respectively. The corresponding regions are marked in
Fig.~~\ref{histo}, and we see that the scaling of $A$ with $\delta_{\rm th}$
is different in these regions.

For large distances, $35 < d/\Mpc < 110$, we find that
$A(\delta_{\rm th},d)$ approaches the self-similar behavior
\begin{equation}
A(\delta_{\rm th},d) = x^{-\beta} A(\delta_{\rm th}\times
x^\alpha, d_0), \label{selfsimilar}
\end{equation}
where $x \equiv d_0/d$. We find that the extrapolation according to this
expression is in good agreement with the direct calculation in the region of
deflections dominated by multiple filament crossing ($0.2^\circ < \delta_{\rm
th} < 1^\circ$ at $d= 110~\Mpc$) if $\alpha = 0.5$ and $\beta = 1$, see
Figure~\ref{histo}. This can be understood as follows. Self-similarity is
consistent with the assumption that the density of deflectors (filaments)
reaches a constant value at large distances. Since MFs are uncorrelated in
different deflectors, multiple crossings should produce a ``random walk'' in
the deflection angle, resulting in deflection angle growing as a square root
of the propagation distance $d$, i.e.~in $\alpha = 0.5$. On the other hand,
the solid angles subtended by individual deflectors decrease as $d^{-2}$, while
their number grows as $d^{3}$, resulting in linear growth of the total area
``shadowed'' by deflectors, i.e.~in $\beta=1$.

We use Eqn.~(\ref{selfsimilar}) to extrapolate $A(\delta_{\rm th},d)$ up to a
distance of $500~\Mpc$. We find that even with such large distances to the
sources, the expected sky coverage of deflections larger than $1^\circ$ does
not exceed 30\%, see Fig.~\ref{histo}.

\begin{figure}[t]
\includegraphics[width=0.47\textwidth]{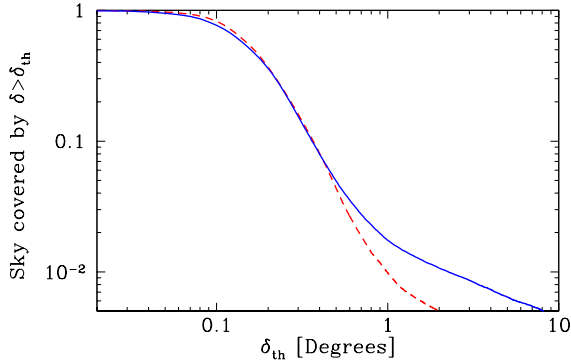}
\caption{Cumulative fraction of the sky with deflection angle larger than
$\delta_{\rm th}$. Two runs with different MF seed are compared for $E
= 4.0\times10^{19}$ eV and propagation distance 110 Mpc. Results of the run
with $B_0 = 10^{-11}$~G are scaled as $A(\delta_{\rm th}/5,d)$ and are shown
by the dashed line.}
\label{fig:fig13}
\end{figure}

It is important to check the sensitivity of the results to the assumed value
of the seed field. We do this in Figure~\ref{fig:fig13}, where we compare the
function $A(\delta_{\rm th})$ obtained for our two simulations with seed
fields producing $B_0 = 10^{-11}$~G and $B_0 =2\times 10^{-12}$~G, 
respectively.
Deflections obtained in the first run are divided by the factor of $5$ and are
shown by the dashed line. Both curves match each other reasonably well in the
region of small deflections, but start do disagree in the region which reflect
deflections in clusters. This difference with respect to a simple linear
scaling should be expected, and arises due to the non-linear amplification and
saturation effects.

As discussed in Section~\ref{results2}, the magnetic field in low density
regions is affected by the coherence and the orientation of the magnetic seed
field.  Projected sheets that appear as filaments of  low density have not
undergone significant non-linear structure formation, therefore their
appearance can be different for different choices of the initial field
orientation. Depending on their orientation with respect to the initial field
vector and the line-of-sight to the observer, the details of the filamentary
structure in deflection maps differ between our two simulations which have
orthogonal initial field directions.  However, Figure~\ref{fig:fig13}
demonstrates that statistically there is no difference between the two
different initial field orientations. Note also that the small filaments are
unconstrained by the 1.2-Jy data imposed on the initial conditions of our
simulations; instead they arise from purely random CDM fluctuations added on
small scales.  Finally, we remark that for a more chaotically oriented initial
seed field the deflections along the filaments, sheets and voids will be
accumulated as a random walk, which compared to the coherent case we have
considered will tend to reduce the deflections. Again, our simulation should
therefore give a robust upper limit.

Our extrapolation of cumulative deflections to distances exceeding the size of
our simulation volume does not take into account the potential effect of a
real unclustered component in the EGMF. This may exist in the voids and low
density regions if the magnetic seed field was produced at high
redshift. While the effect of such a component would not be very significant
in our deflection maps because of the finite size of the simulation, it might
however build up for UHECR produced at cosmological distances, and then give
rise to observable deflections.  Indeed, a uniform field with strength of $2
\times 10^{-12}$ G gives rise to deflections of up to $\approx 1^{\rm o}$ for
protons with energy $4\times 10^{19}~\eV$, over a distance of 500 Mpc.  Note
however that even sizable deflections caused by the coherent field are not
fatal for the identification of UHECR sources, as they can be disentangled
with the use of the method suggested by \citet{Tinyakov:2003nu}.  Moreover, a
seed field uniform over cosmological distances is unlikely. More
realistically, $l_c \ll d$ and the proton trajectory will make a random walk
through the magnetic domains. The overall deflection will then be given by
$$
 \delta \simeq 0.2^o\;
\left(\frac{B_0}{E}\frac{4\times 10^{19}~\eV}{10^{-11}~{\rm
G}}\right) \left(\frac{d}{1~{\rm Gpc}}\right)^\frac {1}{2}
\nonumber\left(\frac{l_c}{1~\Mpc}\right)^\frac {1}{2}.
$$ 
Hence, observable deflections are not produced by the unclustered
component of the IGM if $l_c$ is smaller than a few tens of Mpc.  Note
that such small coherence lengths are expected from most of the
proposed generation mechanisms of seed EGMFs
\citep{Grasso..PhysRep.2000}. The few mechanisms predicting larger
$l_c$ generally give rise to MF which are too weak to produce
observable deflections of UHECR. Furthermore, an unclustered EGMF does
not exist at all if the seed field is generated by a battery powered
by structure formation \citep{Kulsrud..ApJ.1997}. 

\section{Conclusions and Discussion}

In this work we analyzed the results of a constrained cosmological MHD
simulation of the Local Universe embedded within a periodic box of $\simeq
343$ Mpc on a side. The high resolution region centered on the `Milky Way'
covers a sphere with $\simeq 115$ Mpc radius and is resolved by 50 million
dark matter plus the same number of gas particles. The initial conditions
where obtained using constraints from the IRAS 1.2-Jy galaxy survey
complemented by a random realization of a $\Lambda$CDM universe. As a result,
the evolved simulation at $z=0$ reflects the large-scale structure observed in
the real Universe quite well, making it possible to identify most of the
prominent halos and structures found within the simulation with known galaxy
clusters and superclusters of the Local Universe.  This setup removes the
arbitrariness involved in choosing a suitable observer position in ordinary
simulations when trying to estimate how much the observed UHECRs may be
deflected by extra-galactic magnetic fields.

We extracted the most massive clusters within the high resolution
region of the simulation, leading to a set of 16 clusters (volume
limited sample) within a temperature range of $\approx 3-8~{\rm
keV}$. Examining the properties of the magnetic field formed in these
systems, we confirm findings from earlier work. Radial profiles of the
magnetic field strength are similar to that of the gas density in the
outer parts, but the central magnetic field value strongly scales with
the cluster temperature. During cluster formation, the magnetic seed
field is not only amplified by adiabatic compression but also by shear
flows that drive magnetic induction, a process that is ultimately
powered by anisotropic accretion and merging events.  The initial
field geometry is wiped out completely by the violent cluster
formation history, a result that is almost independent of the exact
mechanism for generating the initial magnetic seed field, provided it
is generated early enough, say before $z\sim3$. This makes the
strength of the comoving intensity of the seed field, $B_0$, 
essentially the only relevant free parameter of our model.

We constructed synthetic Faraday rotation measures from the simulated clusters
and compared them to a variety of observational data. We demonstrated that a
comoving seed field of $B_0 = 0.2 \times 10^{-11}$ G reproduces the observed
amplitude and the radial scaling of rotation measurements found in galaxy
clusters very well.  The strong dependence of the cluster magnetic field on
cluster temperature leads to correlations between other measurable physical
quantities. The predictions of our simulations for these correlations can also
be confronted with observations. We demonstrated a good agreement for the 
scaling of the X-ray surface brightness with the rotation measure. Using two
different models for the relativistic electron population, we also
demonstrated that the simulated clusters are able to account for the observed
correlation between radio power and temperature.

In low density regions outside of clusters, such as filaments, sheets and
voids, no significant non-linear structure formation takes place and the
magnetic seed field evolves mainly due to adiabatic compression. While within
these regions large scale flows can still reorient the magnetic field, it will
remain correlated with the geometry of the initial seed field. The detailed
structure of the field in these regions therefore depends strongly on the
properties of the seed, and thereby ultimately on the mechanism which creates
the seed field in the first place.

We calculated the expected deflections of ultra high energy cosmic rays by the
magnetic field in the large-scale structure of the simulated Local
Universe. Given that the simulation reproduces in a realistic way positions,
masses and sky coverage of the prominent clusters and superclusters of the
Local Universe, we have no freedom to choose the observer position. Instead,
we expect that placing the observer at the position of the ``Milky Way'' will
provide a good model for what one expects to see in the real Universe.

We have computed full sky maps of expected deflections for particles with an
energy at the detector of $4\times 10^{19}$eV and $1\times 10^{20}$eV. These
maps show that strong deflections occur only at crossing galaxy
clusters. Filaments, or sheets seen in projection, can lead to some
noticeable deflections, but UHECRs approaching from most areas of the sky will
not suffer significant deflections by the extragalactic magnetic field.  For
UHECRs with arrival energies of $1\times 10^{20}$eV, we also demonstrated
that, if energy losses are included, the deflection will be reduced, as the
traveled distance will decrease (for UHECRs with the same injection energy)
and parts of the distance will be traveled at higher energies. UHECRs with
energies of $4\times 10^{19}$eV could in principal originate from sources up
to a distance of 500 Mpc, which cannot be directly probed by our
simulation. We therefore extrapolated the distribution of deflection angles
for these distances, still finding that most of the sky should be covered by
areas with small deflections.

Using another earlier simulation, where we chose a five times larger initial
seed field that pushes the magnetic field within galaxy clusters towards (or
even slightly above) the upper limit of observationally allowed magnetic field
strength, we verified that even in such a more extreme model a significant
fraction of the sky is covered by deflections below or comparable to 
the resolution of current
experiments. In this second simulation we also used a different initial
direction of the magnetic seed field, allowing us to explicitly verify that
the statistical distribution of deflection angles is insensitive to the
direction of the initial seed field.

A major systematic uncertainty of our model is the structure of the initial
magnetic seed field. However, we argue that our estimates for the typical
deflection angles provide a robust upper limit.  This is because an initially
homogeneous magnetic seed field is expected to lead to the maximum deflection
for a given field strength.  Any tangled component in the seed field will
reduce the deflection and therefore strengthen our final conclusion. Note also
that our modeling neglects contributions to the intra-cluster magnetic field
from local injection processes, related for example to galactic winds or AGNs,
which probably occur also in the late phases of cluster formation. Such an
additional contribution will increase the cluster magnetic field. If important
and accounted for, this would then force us to lower the cosmological seed
field in order to avoid exceeding the observational limits for the field
in clusters. A side effect would then be a reduction of the field and of the
deflection angles in low density regions, again strengthening our final
conclusion.

In summary, we hence conclude that our model provides an upper limit for the
expected deflections of UHE protons by the magnetic field embedded in the
large-scale structure of the Local Universe.  Therefore our main conclusion is
that the deflections of UHE protons with energies larger than $4\times
10^{19}$ eV are not big enough to prevent the pointing of UHECR sources in a
significant fraction of the sky. Charged particle astronomy should be
possible.
Note that for this purpose any deflection of the
UHECRs within the primary sources or within the direct embedding environment
-- like for sources embedded in galaxy clusters -- is irrelevant, as long as
these structures are smaller than the angular resolution of the experiments,
which is true even for galaxy clusters, barring the exception of a very small
number of nearby clusters.

An opposite conclusion was recently reached by
\citet{2004astro.ph..1084S,Sigl:2003ay}, who on the basis of a different MHD
simulation and differing physical assumptions found significant deflections of
UHECRs, large enough to spoil charged particle astronomy.  Their large
deflections are a consequence of the much larger EGMF found in their MHD
simulation, not of their different approach adopted for modeling the
propagation of cosmic ray primaries.  In fact, their field strength decreases
more slowly than in our simulations when going from galaxy clusters to the
field, so that strong MFs extend over a larger volume and give rise to larger
protons deflections.
This difference can be due to several factors. Those which we were able to
identify are listed below: 1) spatial resolution of the numerical 
codes; 2) simulation volume; 3) observer position;   4) physical
assumptions about seed field.

The last issue was studied recently in \cite{Sigl:2004gi}. It was shown that
the use of cosmological seed field, as opposed to Biermann battery process
with assumed subsequent turbulent amplification (which is not really well
defined) does bring the result closer to ours, but cannot account for all
difference.

Regarding the third issue, we note that the average MFs strength and the
mean proton deflection may be significantly increased even if the observer is
placed in the void but nearby a highly magnetized regions, see
e.g. Fig.~\ref{fig:fig1}. Our simulations have clear advantage here. Using
constrained simulations we remove the ambiguity in the choice of the observer
position.

The first and second issues affect the result via different overall
normalization of the original seed field. In both simulations the seed field
is normalized to reproduce MF in the centers of clusters. If for some reason
MF is undershoot in cluster centers in a given simulation, this will lead to
an overestimation of MF in filaments and voids, and vise versa.

In the simulation of \cite{2004astro.ph..1084S,Sigl:2003ay} the largest
clusters have temperature of a few keV, corresponding to a mass significantly
smaller than $10^{15}~M_\odot$, while in our simulation we find clusters with
temperature as large as $8~\keV$ which is consistent with observations. Since
the MF strength in the clusters is correlated with the cluster temperature
\citep{Dolag:2001}, the strength of ICMFs inferred from the RMs in real
clusters cannot be used to normalize the MF in simulations in which the
cluster population is not representative of the real one.

Numerical resolution affect our Lagrangian simulations and the Eulerian
simulations of \citet{2004astro.ph..1084S,Sigl:2003ay} in a similar way. Since
we can resolve the internal structure of galaxy clusters much better than
possible with fixed mesh codes, we are confident that our normalization to the
observational data on clusters is reliable. (We note that the resolution of
both codes is comparable in filaments.) Note also that the analysis of the RMs
of polarized radio sources observed in these systems shows that the coherence
length of ICMFs is typically $\sim 10~\kpc$, comparable to the resolution we
reach in clusters, but ten times smaller than the resolution reached in
\citet{2004astro.ph..1084S,Sigl:2003ay}.

It should be stressed that in our simulations the adiabatic compression and
inductive amplification give roughly equal contributions into the final MF in
clusters, see Fig.~\ref{fig:b_ampli}. This is a non-trivial result which 
has to be verified with future work. 

These discrepancies highlight the significant uncertainties that still
exist in predicting the structure of extragalactic magnetic
fields. Clearly, more work will be required to fully resolve this
issue, which is of significant relevance for the prospects of
forthcoming experiments. Further progress in this will probably be
driven both by refined numerical methods as well as by ever more
demanding comparisons with observational data.  Our present success in
matching the observed correlations between the X-ray surface
brightness and RMs in individual galaxy clusters, as well as the
combined RMs profiles and radio halo luminosity is already highly
encouraging in this respect.

\section*{Acknowledgments}

We would like to thank V.A.~Berezinsky, P. Blasi, G.~Brunetti,
T. Ensslin and G. Sigl for valuable discussions.  The simulations were
carried out at the Computing Center of the Max-Planck Society,
Garching, Germany. Post-processing of the data was carried out on the
IBM-SP4 machine at the ``Centro Interuniversitario del Nord-Est per il
Calcolo Elettronico'' (CINECA, Bologna), with CPU time assigned under
an INAF-CINECA grant. K. Dolag acknowledges support by a Marie Curie
Fellowship of the European Community program 'Human Potential' under
contract number MCFI-2001-01227.


\end{document}